\documentclass[table,11pt,a4]{article}
\usepackage[left=2cm, right=3cm, top=2cm, bottom=2cm]{geometry}
\newcommand*{\RNum}[1]{\uppercase\expandafter{\romannumeral #1\relax}} 
\usepackage{amsmath,amssymb} 
\usepackage[toc,page]{appendix} 
\usepackage{array,multirow,graphicx} 
\usepackage{authblk} 
\usepackage[USenglish]{babel}
\usepackage{blindtext} 
\usepackage{booktabs} 
\usepackage{caption,subcaption} 
\usepackage[font={color=airforceblue,footnotesize,it}]{caption} 
\usepackage{changepage} 
\usepackage[inline]{enumitem} 
\usepackage{float,adjustbox} 
\usepackage[flushleft]{threeparttable} 
\usepackage{footnote}
\usepackage{fullpage}
\usepackage{gensymb} 
\usepackage{graphicx} 
\usepackage{grffile}
\usepackage{ifdraft} 

\usepackage[left]{lineno} 
\usepackage{lscape} 
\usepackage{makecell} 
\usepackage{mathtools,xparse}
\usepackage{nameref,hyperref} 
\usepackage{newunicodechar}
\usepackage{natbib} 
\usepackage[super]{nth} 
\usepackage{pdfpages} 
\usepackage{scrextend} 
\usepackage{setspace} 
\usepackage{soul} 
\usepackage{subcaption} 
\usepackage{svg} 
\usepackage{titlesec} 
\usepackage[colorinlistoftodos]{todonotes} 
\usepackage[utf8]{inputenc}
\usepackage{verbatim} 
\usepackage{wrapfig} 

\usepackage{booktabs}
\usepackage{multirow}
\usepackage{colortbl}

\titleformat*{\section}{\large\bfseries}
\titleformat*{\subsection}{\large\bfseries}
\titleformat*{\paragraph}{\large\bfseries}
\titleformat*{\subparagraph}{\large\bfseries}

\definecolor{airforceblue}{rgb}{0.1725, 0.3490, 0.5255}
\definecolor{risk_red}{rgb}{1.0, 0.0, 0.31} 
\definecolor{risk_orange}{rgb}{1.0, 0.75, 0.0} 
\definecolor{risk_green}{rgb}{0.52, 0.73, 0.4} 
\definecolor{beige}{rgb}{0.91, 0.84, 0.42} 

\hypersetup{
  citebordercolor=white,
  filebordercolor=white,
  urlbordercolor=white,
  linkbordercolor=white
}

\usepackage{subcaption}
\usepackage{caption}
\DeclareCaptionLabelFormat{subfig}{\normalfont(\alph{subfigure})}

\title{Democratising Agricultural Commodity Price Forecasting: The AGRICAF Approach}
\author{Rotem Zelingher\textsuperscript{1} \thanks{rotem.zelingher@wu.ac.at}}
\affil{\textsuperscript{1}Department of Economics, Vienna University of Economics and Business (WU), Vienna, Austria}
\date{}

\begin{document}

\maketitle


\begin{abstract}	
Ensuring food security is a global challenge, particularly in low-income countries where food prices affect access to nutritious food. 
The instability of global agricultural commodity (AC) prices exacerbates food insecurity, with international trade restrictions and market disruptions further complicating the situation. Although online platforms exist for monitoring food prices, there is still a need for accessible, detailed forecasts for non-specialists. 
This paper proposes the Agricultural Commodity Analysis and Forecasts (AGRICAF) methodology, integrating explainable machine learning (XML) and econometric techniques to analyse and forecast global ACs prices up to one year ahead across different horizons.
This innovative integration allows us to model complex interactions while providing clear, interpretable results. 
We demonstrate the utilization of AGRICAF, applying it to three major commodities and explaining how different factors impact prices across months and forecast horizons. 
By facilitating access to reliable forecasts of AC prices, AGRICAF can advance a fairer and sustainable food system.

\end{abstract}

\section{Introduction}

Food prices play a critical role in ensuring food security, directly affecting the accessibility and affordability of nutritious food, especially in low-income countries. The second Sustainable Development Goal (SDG) aims to "end hunger, achieve food security and improved nutrition, and promote sustainable agriculture," emphasising the need to correct and prevent trade restrictions in global agricultural markets \citep{un2015transforming}. International trade in agricultural commodities (AC) can mitigate food insecurity by enabling the transfer of food surpluses from regions of abundance to areas facing shortages, thus maintaining a stable diet and price level throughout the year \citep{costinot2018theusgains, van2017challenges}. Despite global efforts, low-income countries remain vulnerable to price shocks, and price forecasts in the global agricultural markets are often the domain of investors, banks, and large businesses, further complicating efforts to achieve food security \citep{swinnen2012Mixed}.

Achieving a balance in international food distribution and price stability—where surpluses from one region effectively offset deficits in another—is the equilibrium that trade in agricultural commodities seeks to establish. However, this equilibrium faces many challenges. 
Exporting countries often implement protective measures to shield their citizens from food price shocks, particularly during times of crisis. For example, the 2008-2009 food crisis saw significant price increases due to market disruptions, primarily on the supply side. During this period, major exporting countries commonly resorted to export bans or restrictions to mitigate adverse market conditions and protect their populations from food shortages \citep{childs2009factors}. Similarly, the COVID-19 pandemic began with low global supplies and escalated to a global price shock \citep{espitia2020covid}.
While high-income countries can stabilise local prices by enhancing market transparency and providing access to market information, low-income countries often experience sharp price increases during such crises, worsening their food insecurity \citep{hertel2016predicting}. 

The volatility of local food prices in low-income countries is closely tied to global agricultural commodity prices; limited access to market information and resources further destabilises these markets. Farmers and stakeholders in these regions often cannot predict future price movements, negotiate fair prices for their products; nor to adjust their production, storage, or selling strategies for the coming year or months, leaving them vulnerable to price shocks \citep{aker2010information, jensen2007thedigital}. This lack of responsiveness is particularly harmful to small-scale farmers and households whose food expenditures make up a significant proportion of their monthly budgets \citep{usda_food_expenditure}. As a result, they are left highly vulnerable to price fluctuations, with implications for their nutritious status, health, and ability to cope with work and other daily activities. 

The urgency of mitigating this problem led to the establishment of online platforms which provide accessible tracking of the global food prices, followed by "early warning" in case of price shocks. 
Two notable open-access tools, the \href{https://fpma.fao.org/giews/fpmat4/#/dashboard/tool/international}{Food Price Monitoring and Analysis (FPMA)} by FAO and the \href{https://www.foodsecurityportal.org/}{Food Security Portal (FSP)} by IFPRI, which monitor, share, and analyse international and domestic prices of essential food commodities. Both tools include a nowcasting system (Early Warning), reporting on markets with significant volatility. 
The Food Security Portal also supplies information of future prices. 
Another tool is \href{https://dataviz.vam.wfp.org/economic/overview}{The Economic Explorer} by the World Food Programme (WFP). This tool, included in the \href{https://dataviz.vam.wfp.org/}{Vulnerability Analysis and Mapping (VAM) Data Visualisation} Platform, allows users to visualise and download up-to-date commodity price data at the country and market levels.
Yet, these tools, along with the \href{https://www.amis-outlook.org/}{Agricultural Market Information System (AMIS)}, fall short in providing detailed explanations of the data and their influencing factors. This shortcoming ultimately limits the accessibility for non-specialists seeking to assess risks and understand market fundamentals. On top of that, the aforementioned organisations refrain from making and publishing forecasts for AC prices. Despite the availability of these tools, their inherent limitations decrease their broader informative value for non-specialists in formulating effective food security strategies.

To address this gap, we propose the Agricultural Commodity Analysis and Forecasts (AGRICAF) methodology, an innovative approach designed to analyse and forecast global agricultural commodity prices in the short to medium term, up to one year ahead. AGRICAF is distinguished as the first tool to integrate explainable machine learning (XML) with econometric techniques for this purpose. Furthermore, AGRICAF ensures full accessibility to its data sources and insights, breaking down barriers related to budget, education, and language.

AGRICAF’s value lies in its multi-step process. It begins by screening data to ensure consistency, performing retrospective analyses to identify key drivers of price fluctuations, and then applying these insights to forecast prices with high accuracy. This unique combination of methods allows AGRICAF to accurately forecast agricultural commodity prices by recognising patterns in data that signal market changes, even in the absence of real-time information. 
By utilising multiple statistical tests, time series (TS) models, and explainable machine learning (XML) algorithms to capture complex patterns and to interpret relationships between variables, AGRICAF can identify the effects of ongoing events, such as low stocks, trade disruptions, or weather anomalies, and project how these factors will influence prices in the future. 
AGRICAF successfully forecasted wheat price changes a year in advance during the Ukraine-Russia conflict. It identified earlier signs, such as low wheat stocks caused by COVID-19 trade disruptions and extreme weather events during the 2020/2021 season, and used these insights to predict price movements. This ability to detect and account for multiple factors ensures that AGRICAF provides reliable forecasts even during periods of significant market instability or unexpected events. 
As the forecasted period approaches, AGRICAF can adapt its predictions by incorporating updated data. It continuously refines its analysis and forecasts, adjusting for new information, right up to one month before the required due date.
This flexibility ensures that AGRICAF provides accurate and reliable forecasts, even during periods of market volatility or instability.
By employing cross-validation techniques, AGRICAF improves forecast reliability without relying on assumptions or arbitrary parameters. Moreover, the methodology's ability to generate visual and numerical explanations enhances its practical value, allowing users to both trust and understand the forecasts, even without a deep technical background.

AGRICAF has been developed to enhance food security and possesses five key characteristics:
\begin{enumerate*}
\item Accessibility: AGRICAF uses only accessible and regularly updated input data, and the methodology implementation is open-source. 
\item Comprehensiveness: AGRICAF integrates XML and econometric methods to identify both intra- and inter-data interactions.
\item Accuracy: AGRICAF strives to maximise its analysing and forecasting accuracy relative to two market situations—normal market development and extreme market events that cause significant global price movements.
\item Interpretability: AGRICAF provides detailed yet straightforward visual explanations of the drivers behind its results.
\item Practicality: The combination of these principles creates a tool valuable to both researchers and policymakers, as well as farmers, vendors, and other non-specialists, empowering them with actionable insights for informed decision-making in their daily operations and strategies.
\end{enumerate*}

The rational behind AGRICAF suggests that reducing the knowledge gap in global agricultural commodity trade can significantly benefit smallholder farmers, small businesses, and policymakers in low-income countries.
While small farmers may often rely on intermediaries such as governments, agricultural cooperatives, and non-governmental organizations to access market insights, AGRICAF is designed to be accessible and user-friendly, enabling farmers to use it directly if they wish.
For those who would prefer not to use AGRICAF directly, the insights provided by AGRICAF can be communicated to them through intermediaries such as governments, agricultural cooperatives, and non-governmental organisations, who currently lack access to predictive models that offer insights into market speculations and future trends. 
These intermediaries can use AGRICAF’s accurate and interpretable medium-term forecasts of agricultural commodity prices to assists farmers plan their planting and harvesting schedules more effectively, select more profitable land allocation, and negotiate fairer prices. Policymakers can leverage these forecasts to design more effective food security strategies and trade policies. Therefore, by democratising access to critical market information, AGRICAF can contribute to a more equitable and efficient global food system.

This paper presents the methodology of AGRICAF using three exemplary agricultural commodities: maize, soybean, and wheat. These commodities, although variedly characterised in market, social and nutritional aspects, have all been identified as major players in food security: Maize is a crucial agricultural commodity, used as bio-energy, feed, and food, both in developed and  developing countries; Soybean is the most traded tropical grain worldwide and is widely used as a protein source for human and livestock \citep{demaria2020soybean}; Wheat provides about 20\% of the total dietary calories and proteins worldwide, and is fundamental to the diets of both developed and developing regions \citep{shiferaw2013crops}. 

Our results highlight that the optimal choice of forecasting models and explanatory factors varies significantly between commodities, as well as across different months and forecast horizons. In shorter time horizons, fewer variables exert a substantial influence on price forecasts, with financial factors, such as historical prices of the same or related commodities, playing a dominant role. However, as the forecast horizon lengthens, the influence of these financial variables diminishes, and a broader array of factors, particularly those related to agricultural supply, become more prominent. This dynamic shift illustrates the increasing complexity of agricultural commodity pricing over time, as a larger variety of factors come into play with longer forecast horizons. Ultimately, the diversity of influencing factors expands as we predict further into the future, reflecting the intricate and interconnected nature of global agricultural markets.

The structure of the paper is as follows: The next section details the AGRICAF application, including a comprehensive description of its \hyperlink{methods1_cmaaf_stages}{working process}, the \hyperlink{methods2_data}{data}, and the \hyperlink{methods3_models}{models} utilised. Following this, we present the \hyperlink{results1_forecasts}{results} of the price forecasts, accompanied by a detailed explanation using four different dimensions, including a showcase of a recent extreme market shock. Finally, the \hyperlink{discussion}{Discussion} section highlights the main findings within the AGRICAF process by reviewing and analysing the insights hidden behind the numerical results.

\section{Material and Methods}
In this section, we provide a detailed explanation of the AGRICAF methodology used to forecast and explain changes of global agricultural commodity prices. AGRICAF is designed to handle short- to medium-term predictions, utilising publicly available data and integrating advanced econometric and explainable machine learning (XML) techniques. The methodology involves a multi-stage process that includes data collection, retrospective analysis, price forecasting, and interpretation of results.

\renewcommand{\thesubsubsection}{\arabic{subsubsection}} 

This application proposes a comprehensive methodology, AGRICAF, for the short and medium-term analysis and forecasting of monthly global prices of various agricultural commodities. While the overarching workflow remains consistent, distinct inputs are employed for each commodity. This article demonstrates the application of three staple commodities—maize, soybean, and wheat—each subjected to individual investigation. AGRICAF is designed to utilise only publicly available data, and consequently, this paper relies on data from various global, publicly accessible sources. Supplementary Tab.~\ref{data_sources} provides detailed information regarding these data sources.

\subsection{Model output - monthly global price variation}
The process initiated by AGRICAF involves the extraction of global monthly agricultural commodity (AC) price data from the World Bank's commodity market database \citep{worldbankprices}, spanning the time frame from January 1960 to the most recent available report. To mitigate the influence of inflation, all price time series undergo a deflation process, converting them into real 2010 USD values using the corresponding agricultural price index. 

Changes in the global supply of ACs are linked to the local Market Year of each area, as the timing of harvesting and market availability in different regions directly impacts the overall global supply chain and commodity prices \citep{fasusda}.
Accordingly, the dependent variable in the analysis, $p_{m,y}$, was defined as the proportion (percent) of price change relative to the corresponding month ($m$) of the preceding year ($y-1$), such as $y=1,2,…,Y$.
Further details about the calculation process are in Appendix ~\ref{appendix_deflation}.

Fig.~\ref{obs_price} (right) shows the global monthly nominal prices of three representative AC's during the studied period.
The price trends of these commodities tend to be synchronised, with wheat showing more radical price fluctuations compared to the other two. Wheat, being a non-energy crop, experiences more volatile changes, especially during global supply chain disruptions \citep{headey2010reflections}. 
On the left, the figure presents the consumer food price index across different regions: Northern America and Northern Europe (high-income economies), and South America and Western Africa (low- and middle-income economies). While food prices have risen globally, the increase is significantly more pronounced in lower- and middle- income regions, and food prices seem to be less stable. 

\begin{figure}[H]
    \centering
        \includegraphics[width=\textwidth]{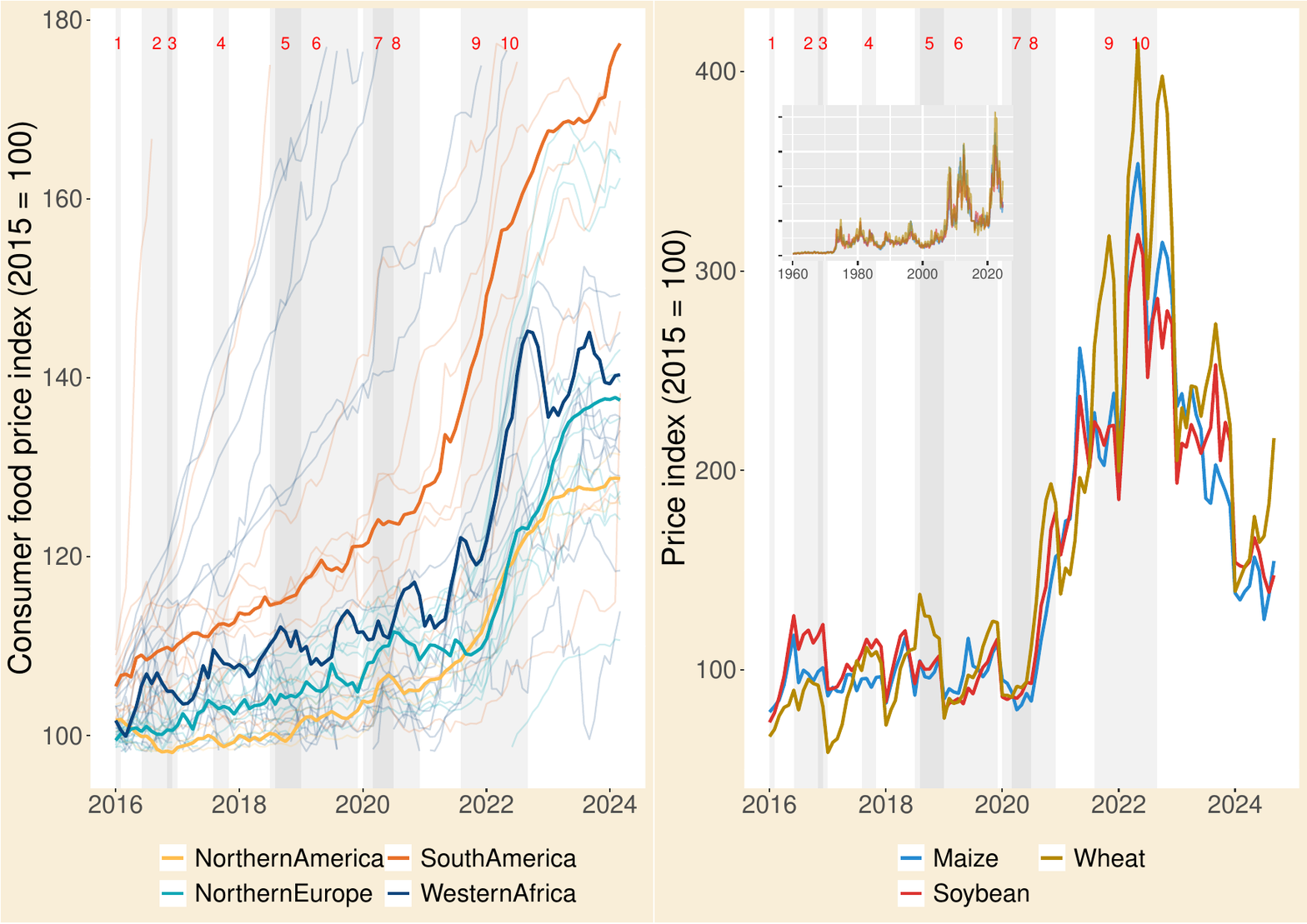}
    \caption{Comparison of Consumer Food Price Index and Agricultural Commodity Prices Over Time.
    On the left, consumer food price indices across several FAO regions. The thick lines represent the median values for each group, with the basis year set at 2015=100. 
    On the right, price indices of three key agricultural commodities: maize, soybean, and wheat, also based on the 2015=100 benchmark. Prices are shown since 2016. The small inner plot shows the complete time series of the same prices since 1960.
    The vertical grey areas signifying significant global events: 
    (1) 2015 El Niño; 
    (2) 2016 Brexit Vote; 
    (3) 2016 Indian Cash Crisis; 
    (4) 2017 Hurricanes Harvey, Irma, and Maria; 
    (5) US-China Trade War; 
    (6) 2019 African Swine Fever; 
    (7) 2020 Locust Swarms in East Africa;
    (8) COVID-19 Pandemic;
    (9)2021 Fertiliser Price Surge;
    (10) Ukraine Russian War.}
    
    \label{obs_price}
\end{figure}

\subsection{Model predictors}
AGRICAF is designed to handle a wide range of variables, provided that the data series is sufficiently long to yield reliable and robust models. The methodology is engineered to import, process, and screen data from various public sources through \hyperlink{methods1_cmaaf_stages}{several stages}. 
This article illustrates the application by using datasets from three sources, as follows: 

\cite{worldbankprices} is a source for global monthly prices of several AC's and price indices. Prices of AC's are treated similarly as the output variable, before being examined, organised and filtered at Stage 1. The handling of price indices, such as energy and fertilisers, follows a similar deflation and conversion procedure, as $p_{m,y}$, to be defined as $x_{k,m,y}^{lag}$.
These variables serve as predictors using different lag effects, such as $h$=1,...,12 months.

\cite{faostat} provides information about annual production, harvested area and yield in country and regional levels, as defined by FAO (transformed into AGRICAF from \href{https://fenixservices.fao.org/faostat/static/documents/Default_coding_and_flags.pdf}{FAO coding system}).
Our stocks data is imported from \cite{fasusda} in country scale, and transformed into AGRICAF from \href{https://apps.fas.usda.gov/psdonline/app/index.html#/app/downloads}{FIPS coding system}. To arrive in regional scales, we use \href{https://fas.usda.gov/regions}{another division}, as used by the USDA. Historical changes are handled by the application to adapt the data to today's political situation. To illustrate, a representative regional cluster is European Union, spanning 32 reported countries including historic territories, such has Former Czechoslovakia, EU15 and previously divided Germany. Also, we consider the UK as part of the EU group.


Here, $x_{k,y}$ indicates the relative change within a specific year and area. Annual production, yield, and stock changes were computed at both national and regional scales.
After undergoing examination, analysis, and screening in \hyperlink{04_calculation}{Stage 1}, they were incorporated into four distinct sets of predictors due to their significant correlations. It is noteworthy that the forecasted price change for a given year and month was based on data available prior to that month. To accomplish this, we utilised local crop calendars \citep{fasusda} to ascertain the local trading year across all considered countries and regions relative to the respective agricultural commodity. 

In total, each monthly price forecast, denoted as $\hat{p}_{y,m}$, is generated using four different datasets, each considering 12 possible forecast horizons. These datasets consist of the following types of explanatory variables:
\begin{enumerate}
    \item \textbf{Variables of monthly frequency}: These variables, represented as $X_{m,y,k}^{lag}$, vary depending on the specific characteristics of the commodity being analysed. Examples include monthly prices, trade data, etc.
    \item \textbf{Variables of yearly frequency}: These variables capture supply-side conditions, which have crucial impact on commodity prices. Here the $m$ index is fixed, such as $X_{\bar{m},y,k}$.
    \begin{itemize}
        \item \textbf{Regional variables}: Two datasets, each with up to 19 variables for regional production or yield, and up to 15 variables for regional stocks. This results in a maximum of 34 yearly variables per dataset.
        \item \textbf{Local variables}: Two datasets, each consists of up to 21 variables representing production or yield from the top-producing countries, and up to 21 variables for stocks in countries with the highest stock levels globally. This sums up to a maximum of 42 yearly variables per dataset. 
    \end{itemize}
    These variables are defined as $X_{\bar{m},y,k}$
\end{enumerate}
Where $y$ and $m$ form a time $t$, each dependent variable can be defined as

\begin{equation}
  p_{m,y}=f(X_{m,y,k}^{lag},X_{\bar{m},y,k})=f(x_{m,1,k}^{lag},x_{m,2,k}^{lag},x_{m,Y,k},...,x_{\bar{m},1,k},…,x_{\bar{m},Y,K})  
\end{equation}

for $k$=1, …, $K$ and $y=1,2,…,Y$.

 \label{methods2_data}

\subsection{The Models} \label{models}
AGRICAF entertains the models as in Tab.~\ref{models_tab}. 
It relies on econometric and XML techniques that provide interpretable short and medium-term forecasts, that can be used by a large audience. Recent developments in technology and research approaches have accelerated the application of statistical and ML algorithms. These models solve complex problems using relatively simple methods while providing predictive results of considerable accuracy, even when compared to particularly advanced models \citep{lobell2010ontheuse, storm2019machine}. 
Tab.~\ref{models_tab} provides a list of these models, along with the stages in which they are used. 

\begin{table}
\centering
\begin{tabular}{llccccc}
\toprule
\multicolumn{1}{c}{} & \multicolumn{4}{c}{Stage} & \multicolumn{2}{c}{Source} \\
\cmidrule(l{3pt}r{3pt}){2-5} \cmidrule(l{3pt}r{3pt}){6-7}
Model & Model Description & I & II & III & Type & Package\\
\midrule
\cellcolor{gray!10}{ARIMA} & \cellcolor{gray!10}{AR Integrated Moving Average} & \cellcolor{gray!10}{} & \cellcolor{gray!10}{} & \cellcolor{gray!10}{\checkmark} & \cellcolor{gray!10}{TS} & \cellcolor{gray!10}{\citep{package_forecast}}\\
CART & Classification and Regression Trees &  & \checkmark & \checkmark & XML & \citep{rpart_package}\\
\cellcolor{gray!10}{GAM} & \cellcolor{gray!10}{Generalised Additive Model} & \cellcolor{gray!10}{} & \cellcolor{gray!10}{\checkmark} & \cellcolor{gray!10}{\checkmark} & \cellcolor{gray!10}{XML} & \cellcolor{gray!10}{\citep{gam_package}}\\
GBM & Gradient Boosting Machine &  & \checkmark & \checkmark & XML & \citep{gbm}\\
\cellcolor{gray!10}{LM} & \cellcolor{gray!10}{(Multivariate) Linear model} & \cellcolor{gray!10}{} & \cellcolor{gray!10}{\checkmark} & \cellcolor{gray!10}{\checkmark} & \cellcolor{gray!10}{XML} & \cellcolor{gray!10}{\citep{broom}}\\
\addlinespace
RF & Random Forests &  & \checkmark & \checkmark & XML & \citep{randomforest_package}\\
\cellcolor{gray!10}{TBATS} & \cellcolor{gray!10}{Advanced time series*} & \cellcolor{gray!10}{} & \cellcolor{gray!10}{} & \cellcolor{gray!10}{\checkmark} & \cellcolor{gray!10}{TS} & \cellcolor{gray!10}{\citep{package_forecast}}\\
VAR & Vector AutoRegressive & \checkmark &  & \checkmark & TS & \citep{vars_package}\\
\cellcolor{gray!10}{XGBoost} & \cellcolor{gray!10}{Extreme Gradient Boosting} & \cellcolor{gray!10}{} & \cellcolor{gray!10}{} & \cellcolor{gray!10}{\checkmark} & \cellcolor{gray!10}{XML} & \cellcolor{gray!10}{\citep{xgboost}}\\
\bottomrule
\end{tabular}
    \begin{tablenotes}
      \small
      \item * TBATS = Exponential smoothing state space model with Box-Cox transformation, ARMA errors, Trend and Seasonal components
    \end{tablenotes}
\caption{Models used in AGRICAF, including the stages they are included in, type of the model, and the package source.}
\label{models_tab}
\end{table}

AGRICAF involves the forecasting accuracy comparison of four XML decision-tree-based models (Group 1), namely CART, RF, GBM, and XGBoost with a tree booster; and three types of linear models (Group 2), i.e., LM, GAM and XGBoost with a linear booster. These models predict annual price changes $p_{m,y}^{h}$ in 12 possible monthly horizons $h$. For each month $m$ and forecast horizon $h$ the function of $x_{k,y}$, $k$=1,..., $K$ utilise a set of predictors, as chosen through an extensive analysis process. The models are tuned through a grid search to identify the optimal hyperparameter setting.	

The application of the Group 2 models involves the removal of multicollinearities from the dataset (see further explanation in App.~\ref{multicollinearities}). This is done separately for each model, as follows:
\begin{itemize}
    \item \textbf{LM}: For each independent variable, we start by detecting aliased (highly collinear) variables from the complete dataset. We then finalise the set of predictors using a stepwise selection based on the \cite{akaike1974aic} information criterion (AIC). 
    \item \textbf{GAM}: When using GAMs, we apply smoothing penalties to prevent overfitting.
    \item \textbf{XGBoost with a linear booster}: In each iteration, we detect and remove multicollinearity from the training set using the \texttt{findCorrelation} function from the caret package \citep{caret_package}, with 0.6 or 0.9 correlation cut-off.
\end{itemize}

 \label{models}

\section{Calculation}
\renewcommand\thesubsubsection{\textbf{S.\arabic{subsubsection}.}} 
AGRICAF combines econometric and explainable machine learning (XML) methods to forecast AC prices over a one to twelve months time horizon.
The methodology offers detailed visual explanations to interpret the results and underlying mechanisms comprehensively. By integrating different econometric and XML methods, AGRICAF evaluates the combined impacts of various potential variables. 
Different cross-validation techniques are employed to avoid prior research assumptions and realistically capture complex relationships.

\begin{figure}
    \centering
    \includegraphics[width=0.5\textwidth]{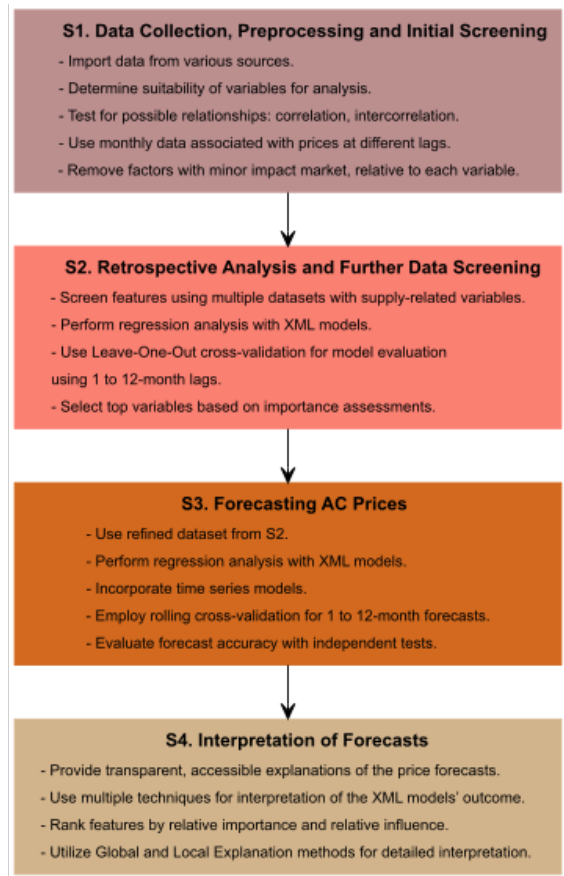}
    \caption[]{AGRICAF's process from the basic problem to the final result, including the working stages}
    \label{cmaaf_chart}
\end{figure}

The workflow of our methodology consists of four main stages, as shown in Fig.~\ref{cmaaf_chart}. 
The methodology starts with stationarity and causality tests to assess variable suitability for analysis, using monthly data linked to prices at different lags, and country/region-level data on production, yield, and stocks from the top 21 countries.
Secondly, AGRICAF performs a retrospective analysis, considering all the variables filtered in Stage 1. Trained on multiple datasets containing supply-related explanatory variables from various geographic scales, this stage provides a secondary screening of features.
Thirdly, AGRICAF reduces the number of impactful factors to finalise the price prediction. 
Lastly, it offers detailed visual and numerical interpretations of the results and learning process, making them accessible to any user. AGRICAF can be easily trained with publicly available data. It is adaptable and applicable to various agricultural commodity markets, regardless of budget, language, or other constraints.

Our methodology relies on econometric and explainable ML techniques \citep{lundberg2017shap,molnar2022} to provide trustworthy short- and medium-term forecasts of agricultural commodity (AC) price changes. AGRICAF aims to find an optimal data-model-training combination to forecast AC prices using both recent and traditional techniques. Currently, AGRICAF combines four ML techniques and five econometric models, detailed in Table~\ref{models_tab}. The working process for each AC price is divided into four stages (S), as follows:

\subsubsection{Data Collection, Examination, Modification, Organisation and Filtration} 

Stage 1 involves the collection of time series data on prices and various potential explanatory variables for the global price of a chosen AC and their verification. Data is modified as detailed in the \hyperlink{methods2_data}{Data} section, prior to be examined and organised coherently. Finally, a cautious screening creates several a baseline datasets, each contains the same dependent variable but different number and type of indicators.  

Initially, Augmented Dickey-Fuller (ADF) tests assess the stationarity of the dependent variable $P$ at a 5\% significance level. We analyse only cases where $P$ is stationary to ensure consistency. The ADF test is conducted in the R statistical environment \citep{rcore} using the "ucra" package \citep{urca_package}. 
For any stationary $P$ time series, potential factors are divided into default ($x_{k,t}^d$) and additional case-specific variables ($x_{k,t}^a$), as described in the \hyperlink{methods2_data}{Data} section.

Monthly data is associated with prices at different lags, depending on the forecasting horizon. Country and region level data are separately used for production or yield and beginning stocks. For each variable we take data for the highest 21 countries, calculated as sum across the whole period.

\subsubsection{Retrospective Analysis and Second Data Filtration}

Stage 2 conducts a retrospective analysis to identify variables driving price shifts in global AC prices through annual relative changes. This analysis involves regression for relative annual changes using different XML models (see Tab.~\ref{models_tab}). The analysis follows a supervised learning procedure \citep{mohri2018foundations}, dividing the complete dataset into a training set of $T$ years and a single year as a testing set, $\hat{t}^{id}$, where $T = Y$-1. Models are trained on the $T$ years to estimate the monthly price function of the general form $p_m=f(x_{k,y})_{m,lag}$, where $x_{k,y}$ are collections of independent variables, and the analysed $lag$ and month $m$ are fixed. Each iteration involves different combinations of hyperparameters, explained in the \hyperlink{methods3_models}{Models section}, from which the best performing forecast is chosen. This process is repeated $Y$ times - once for every observation.
Model performance is evaluated using six error matrices, as detailed in Tab.~\ref{tab_evaluation}.

\begin{table}[h]
    \centering
    \begin{tabular}{p{2cm} p{6cm}p{6cm}}
        \toprule
        \rowcolor{gray!6}
        \textbf{Abb} & \textbf{Metric} & \textbf{Formula} \\
        \midrule
        MAE & Mean Absolute Error & $\frac{1}{n} \sum_{i=1}^{n} \left| observed_i - predicted_i \right|$ \\
        MAD & Meadian Absolute Deviation & $\text{median}(|observed_i - predicted_i|)$ \\
        MAPE & Mean Absolute Percentage Error & $\frac{1}{n} \sum_{i=1}^{n} \left| \frac{observed_i - predicted_i}{observed_i} \right| \times 100$ \\
        MSE & Mean Squared Error & $\frac{1}{n} \sum_{i=1}^{n} \left( observed_i - predicted_i \right)^2$ \\
        RMSE & Root Mean Square Error & $\sqrt{\frac{1}{n} \sum_{i=1}^{n} \left( observed_i - predicted_i \right)^2}$ \\
        RA & Relative Advantage & $1-\frac{RMSE}{std(observed_i)}$ \\
        \bottomrule
    \end{tabular}
    \caption{Regression model evaluation metrics used in AGRICAF}
    \label{tab_evaluation}
\end{table}

To evaluate the models' precision and adjust ML settings without causing over- or under-fitting, we employ Leave-One-Out cross-validation (LOOCV). For each set of input, and for every month $m$ and forecast horizon $h$, the two models with the lowest sum of errors are selected for the next factor screening. Variable importance is cautiously assessed through each iteration relative to the chosen model. Results are scaled and normalised, leading to the calculation of the average importance for each variable. Final screening retains up to 19 highest-ranked variables, enhancing analysis robustness and efficiency.

\subsubsection{Forecast the AC prices}
In Stage 3, the refined dataset from Stage 2 serves as the input for the forecasting models outlined in Tab.~\ref{models_tab}. The dataset is split into a training set comprising the first $T$ years of observations (with 44$ \leq T$) and a testing set including the observation following year $T$.
AGRICAF forecasts global AC prices for 1 to 12 months beyond the final price observation, using iterations of one-step-ahead predictions. 
To maximise forecast accuracy, a rolling cross-validation approach \citep{hyndman2018forecasting} is employed, adding one year of data to the training set in each iteration. This methodology enables forecasting AC price changes using data up to the most recent year. By selecting $T = 44$ as the shortest training period, we ensure a comprehensive training set that covers a broad range of historical data, providing a solid foundation for model training \citep{zelingher2022maize}.
As iterations advance, the training set expands to include observations such as $T$=45, 46, and so on, effectively capturing time-based patterns and variations in the global ACs market. After training, the algorithm produces a forecast, with $T+1 = \hat{y}^f$ as a one-year testing set, and the price to forecast for $\hat{p}_{(\hat{y}^f)}$. Iterations continue until the model predicts prices up to one year into the future.

Additionally to the explainable machine learning (XML) models, our approach incorporates three distinct time series models: Vector Autoregression (VAR), Autoregressive Integrated Moving Average (ARIMA), and Trigonometric seasonality, Box-Cox transformation, ARMA errors, Trend and Seasonal components (TBATS). VAR and ARIMA models are trained on monthly multivariate time series; and TBATS, a univariate time series model, is trained solely on its own historical data.

These models perform an $H$-step recursive forecast \citep{cheng2006multi-step}, with each forecast consisting of $H-1$ inner iterations before arriving at the final prediction. For $1 \leq H \leq 12$, $H$ represents the forecast horizon $h$ in months. Models are trained with different hyperparameter combinations, resulting in various model variants. This approach leverages long time series data (12 observations per year) and can be trained up to current prices. Consequently, it offers an alternative that potentially provides more accurate and comprehensive predictions of AC price changes, incorporating recent market trends. As in the XML algorithms, every inner-iteration is a one-step ahead prediction, only that for every $h<H$, the predicted price joins the input data for the following forecast, along with the new forecasting error to create a new input set of $T^ts$ observations and $h$-1 predicted values. As the forecasting errors accumulate, TS model's performance decreases with any additional iteration.
A minimum training set of 150 observations is used.

While TBATS executes an $H$-step recursive forecast using a uni-column training set, it forecasts $\hat{p}_{(\hat{y}^ts)}$ by capturing complex seasonal patterns using trigonometric representation, without considering external variables. ARIMA can incorporate external variables in its predictions and is particularly effective for data with clear trends or seasonal patterns. VAR extends ARIMA by modelling multiple interdependent time series, capturing linear interdependencies among multiple variables. Together, these models enhance AGRICAF's forecasting robustness and reliability.

To evaluate the forecast accuracy of the models, independent tests similar to those in Stage 2 are conducted. While the XML models use datasets with one observation per year, the time series models capture 12 observations per year, resulting in significantly longer time series. This provides the time series models with a larger training set, which improves their ability to detect patterns and trends over time. However, this increased frequency of observations comes at the cost of using fewer explanatory variables, as time series models skip the supply-side predictors. This trade-off between dataset length and the number of variables influences the models’ predictive performance and the type of insights they provide.

\subsubsection{Interpret the forecasts for global AC prices}

Enhancing the interpretability of model outputs is crucial, especially for non-specialist users \citep{spavound2022trustworthy}. Stage 4 incorporates various model-agnostic techniques inspired by \citet{molnar2022}, providing transparent and accessible explanations of model outcomes, fostering greater user confidence. Although the time series models utilised in AGRICAF lack interpretability, every forecast is reported using at least one explainable model. Selected explainable models undergo comprehensive retraining using the entire dataset, followed by implementing global and local agnostic methods to interpret forecasting results and their underlying mechanisms.

In this final stage, AGRICAF focuses on the outcomes generated by a single explainable model, utilising distinct evaluation criteria. Each model's performance is assessed monthly, relative to a distinctive forecast horizon. This paper presents interpretation figures using two criteria: Mean Absolute Error (MAE) and Relative Advantage (RA). MAE indicates forecast accuracy, with lower values representing more accurate predictions. 
RA is a normalised version of RMSE, calculated as 1 minus the RMSE divided by the data's standard deviation, similar to a standard model skill score. It measures the model's performance relative to a baseline forecast, such as assuming future price changes equal to the mean of past observations (defined as a naïve forecast). As such, RA values above 0 indicate superior model performance compared to a a naïve forecast, and higher values representing more accurate forecasts.

The feature ranking used in Stage 4 is twofold. The first type consists of feature importance methods from Stages 1 and 2, aimed at identifying variables that affect model prediction accuracy. The second type assesses the marginal influence of each feature by comparing its median absolute Shapley value \citep{shapley_1952} to the total median Shapley values of all features. After forecasting the price, these methods evaluate each feature's contribution to prediction accuracy for a given month and forecast horizon.

Interpretation starts with Global Agnostic methods, which can be comparative, reflecting the relative influence of each variable on the predicted price, or descriptive, outlining the average behaviour of a chosen variable over the time series (TS). We report the median role of features within the model in a visually accessible manner.

Local Agnostic methods explain a single model outcome, providing information about market changes driving specific price changes. This involves understanding each predictor's impact on individual price change predictions, combining Shapley values with relative importance used previously. This combination offers insights into each feature's contribution to predicted price variations, identifying potential price shock origins. Visualising these values against predictor values assesses price shock risks relative to regional production variations. 
Tab.~\ref{tab:xml_methods} presents the explanation techniques included in AGRICAF.

\begin{table}
\centering
\begin{tabular}{>{\raggedright\arraybackslash}p{3cm}>{\raggedright\arraybackslash}p{12.5cm}}
\toprule
Method & Mission\\
\midrule
\cellcolor{gray!6}{Coefficients} & \cellcolor{gray!6}{Show expected $p_{m,t}$ changes following a 1\% change in $x_{k,m}$}\\
Residual Analysis & Assess the goodness of fit and identify patterns in residuals\\
\cellcolor{gray!6}{Decision Tree} & \cellcolor{gray!6}{Visualise the relative impact of $x_{k,m}$ on the computing process}\\
Feature Importance & Reflect the contribution of the inclusion of a variable to the overall performance of the model\\
\cellcolor{gray!6}{Global Surrogate} & \cellcolor{gray!6}{Explain a black box model by using its outcome as the training input of a non-black-box model.}\\
\addlinespace
Shapley based summary plot & A combination of Importance value (relative importance) and the global version of the traditional Shapley Decomposition\\
\cellcolor{gray!6}{PDP} & \cellcolor{gray!6}{Similar to PDP, with the added value of combining the variance of $x_{k}$ and presenting each $t$ as a separate point}\\
Shapley Value & The impact of $x_{k,t}$ on a specific price change anomaly.\\
\cellcolor{gray!6}{Features Interactions} & \cellcolor{gray!6}{The alteration in $p_{m,y}^{h}$ that happens when the features are varied, taking into account the individual relative influence.}\\
LIME & Local Interpretable Model-agnostic Explanations: Understand predictions for specific instances\\
\bottomrule
\end{tabular}
\caption{Overview of the model-explanation techniques included in AGRICAF, including method’s group and mission.}
\label{tab:xml_methods}
\end{table}\label{methods1_cmaaf_stages}

\section{Results}
AGRICAF combines econometric and machine learning (ML) methods to forecast agricultural commodity (AC) global price changes for one to twelve-month horizons. The methodology integrates eight econometric and ML techniques, jointly considering over 100 potential variables and allowing for the inclusion of additional explanatory factors as needed. Employing different cross-validation techniques, AGRICAF brings the need for priori assumptions to the possible minimum, realistically capturing complex relationships in the data. 
The workflow consists of four main stages, starting with identifying suitable explanatory variables, continuing with retrospective analysis and variable reduction, until performing the final price prediction. 
Following prediction, AGRICAF offers detailed interpretations of the forecast, explaining the role of each feature in price changes through four levels of explanation. 
This approach ensures comprehension across diverse audiences.
To give an example for possible use of AGRICAF, this article shows the application to three staple commodities - maize, soybean, and wheat - each individually investigated. The data used includes global monthly prices and price indices of commodities \citep{worldbankprices}, annual production and yield at country and regional levels \citep{faostat}, and country-scale stocks data \citep{fasusda}.

\subsection{AC Prices are Accurately Predictable with Publicly Available Indicators}
\begin{figure}[H]
    \centering
    \begin{subfigure}[b]{0.49\textwidth}
        \includegraphics[width=\textwidth]{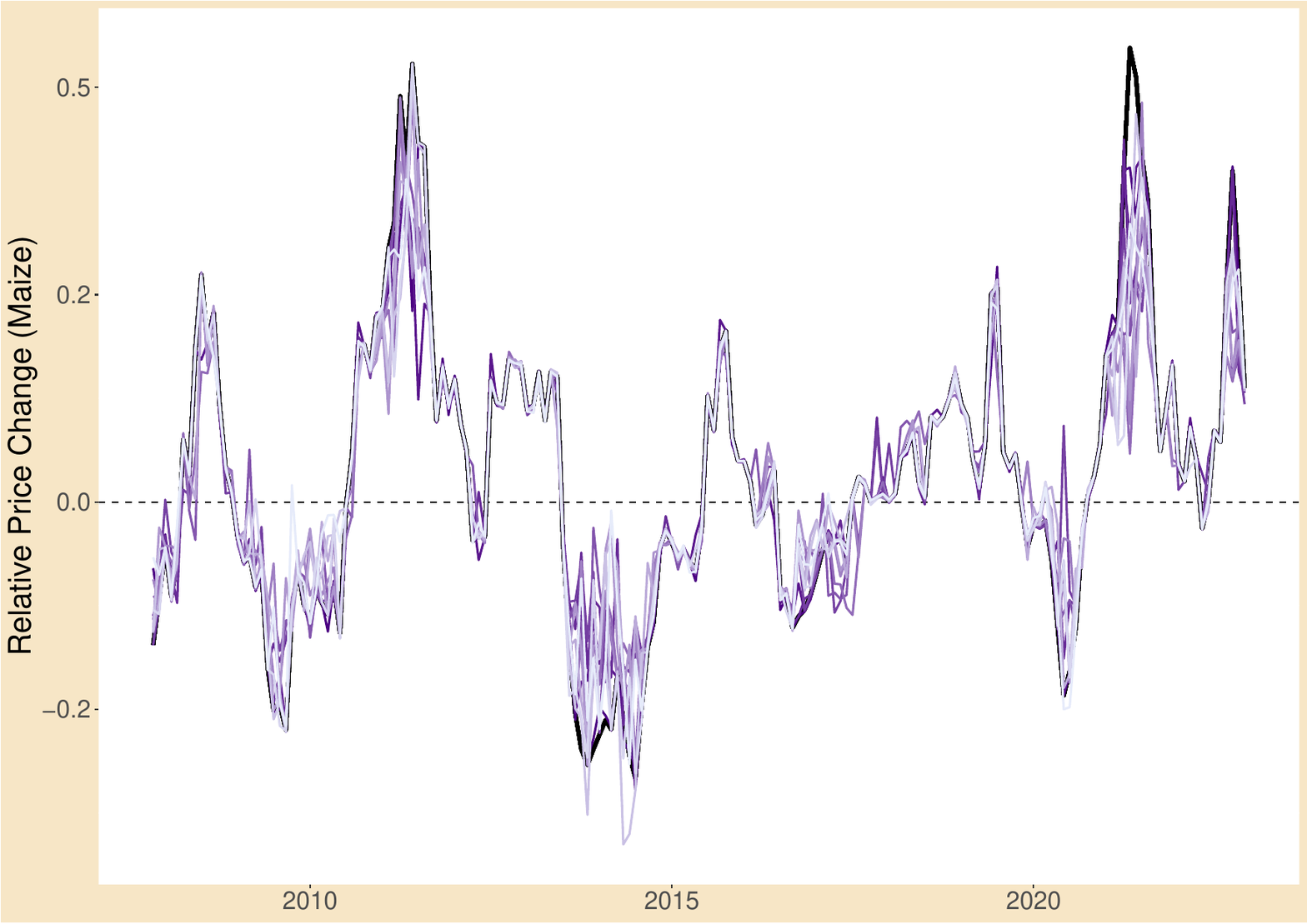}
        \caption{Maize}
        \label{fig:maize}
    \end{subfigure}
    \hfill
    \begin{subfigure}[b]{0.49\textwidth}
        \includegraphics[width=\textwidth]{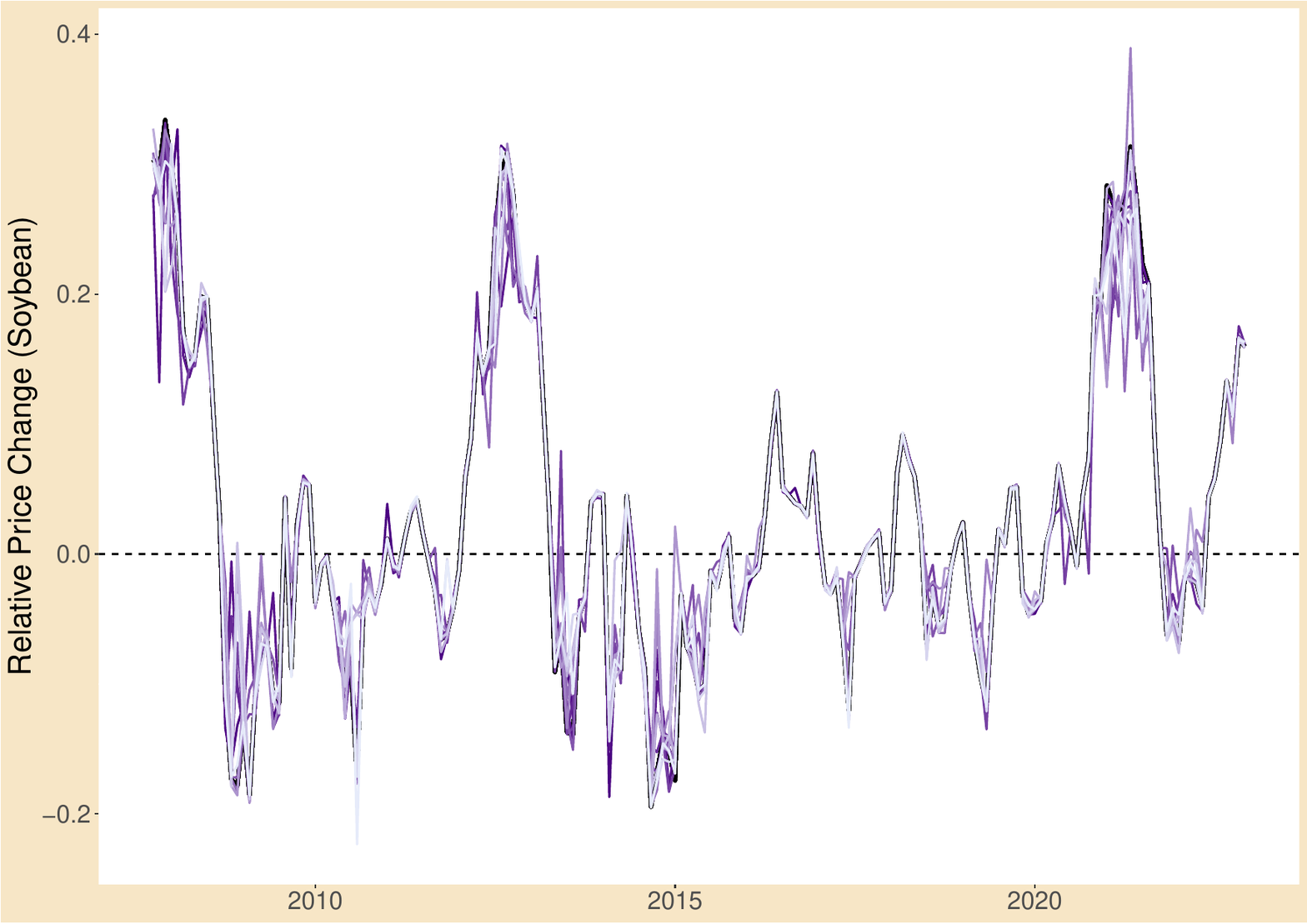}
        \caption{Soybean}
        \label{fig:soybean}
    \end{subfigure}
    
    \begin{subfigure}[b]{0.49\textwidth}
        \includegraphics[width=\textwidth]{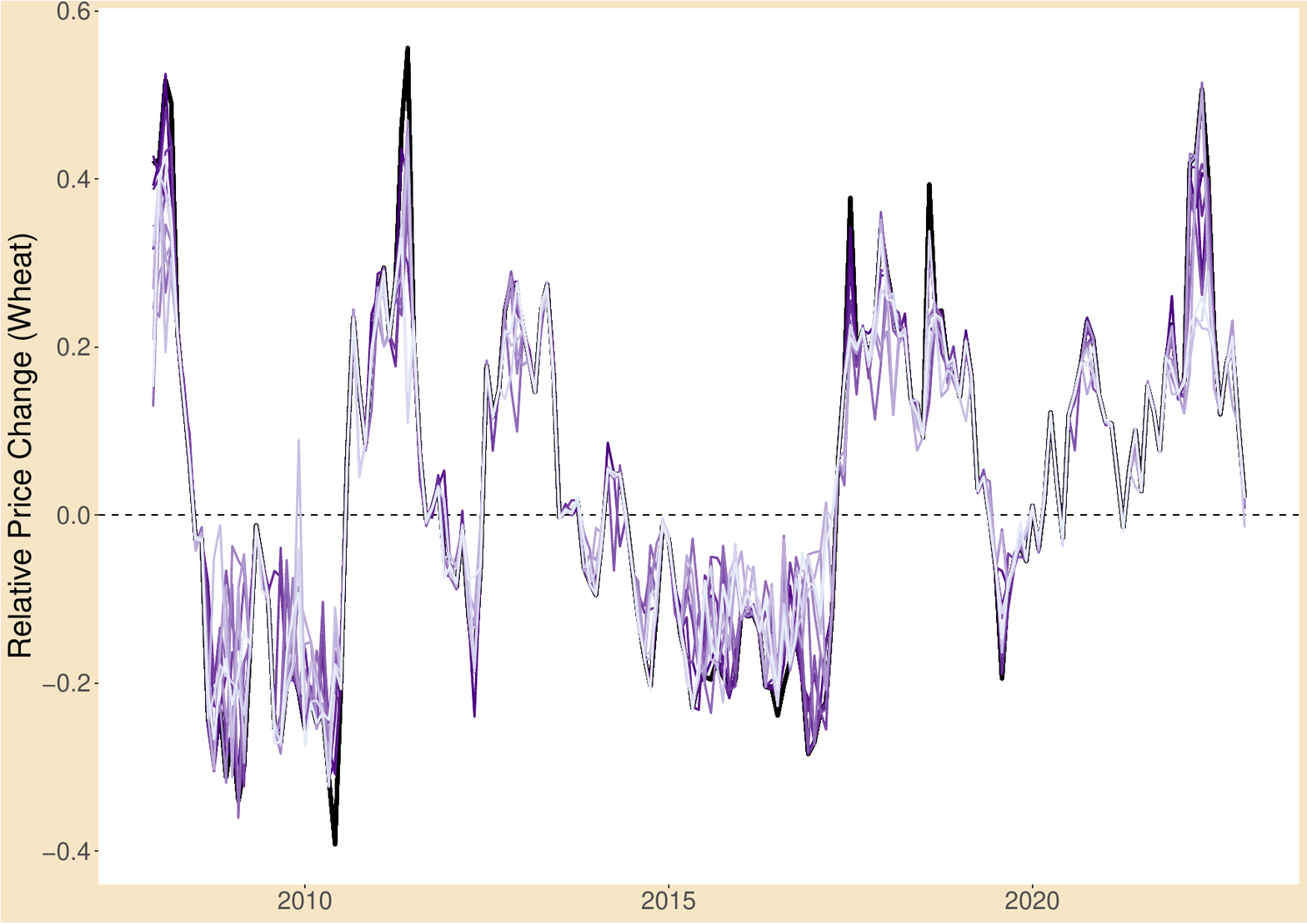}
        \caption{Wheat}
        \label{fig:wheat}
    \end{subfigure}
    \hfill
    \begin{subfigure}[b]{0.49\textwidth}
        \includegraphics[width=\textwidth]{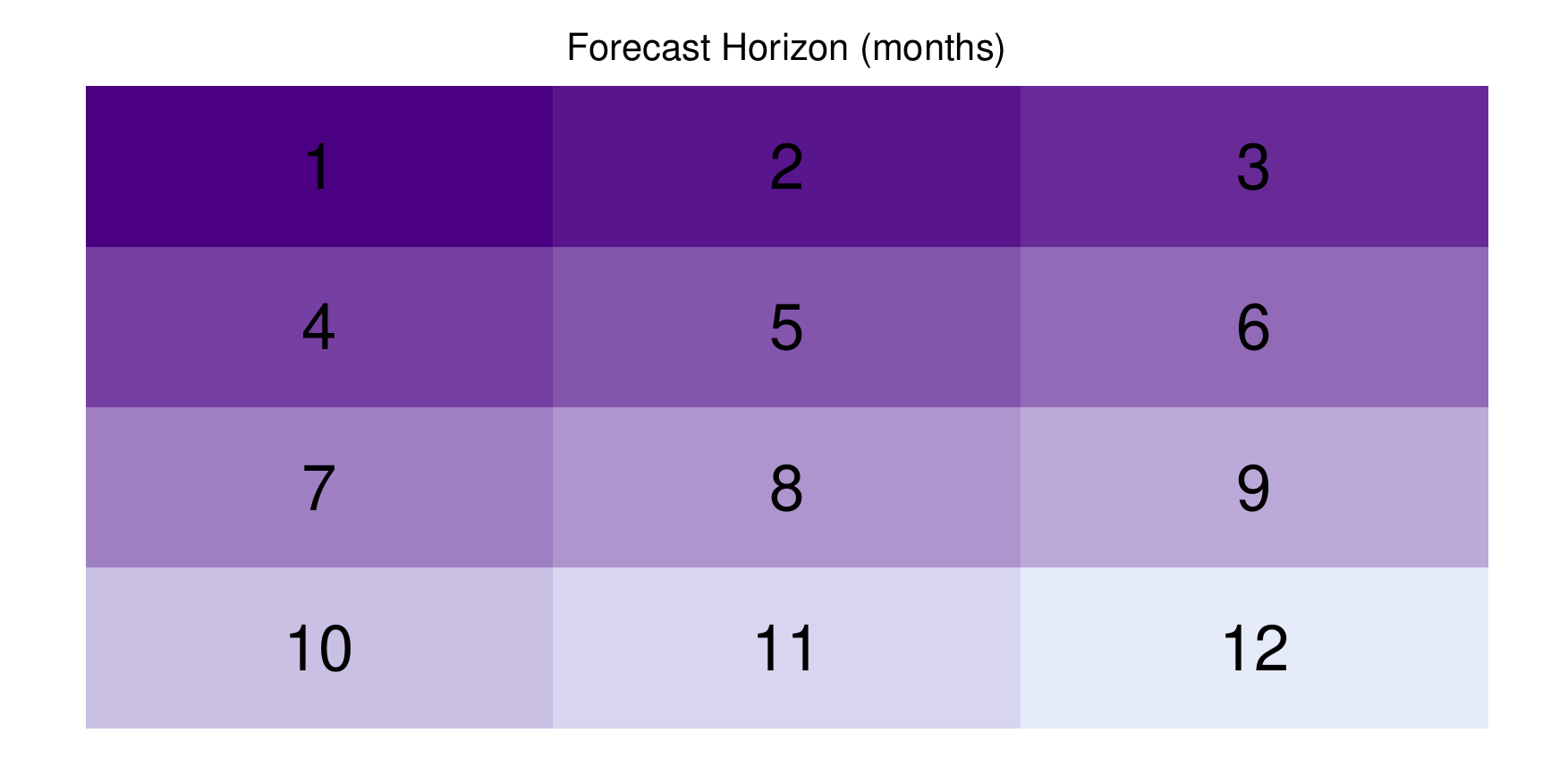}
        \label{fig:legend}
    \end{subfigure}
    \hfill
    \caption{Comparative analysis of observed and predicted relative annual price changes (monthly prices, December 2007-2024). The black line represents the historical price changes for maize (a), soybean (b), and wheat (c). The purple lines indicate the predictions from the highest performing model, optimised for MAE minimisation, as shown in the legend at the bottom right, for horizons up to 12 months. Detailed graphics are in Fig. ~\ref{pred_obs_maize}, ~\ref{pred_obs_soybean}, and ~\ref{pred_obs_wheat}.
    Detailed list of events associated with large forecast errors of each commodity is in Tab. ~\ref{tab_extremes_explained}.} 
    \label{obs_pred}
\end{figure}

Fig.~\ref{obs_pred} presents a comparative analysis of observed and predicted (in sample) global price changes for three distinct agricultural commodities in focus — maize (a), soybean (b), and wheat (c) — relative to twelve forecasting horizons. The temporal dynamics of these commodities are represented over the years 2007 to 2024.
For each commodity, the observed relative price change is shown in the black line. The coloured lines show the forecasts obtained by the model with highest accuracy, relative to forecast horizons. 
Numeric results of the average prediction accuracy are given in Appendix ~\ref{appendix_results} in the form of two error matrices. 

As shown in Fig.~\ref{obs_pred}, the wheat (St.Dev = 0.18) sector tends to higher fluctuations compared to maize (St.Dev = 0.16) and soybean (St.Dev = 0.13). Fig.~\ref{obs_pred} also visualise that, in terms of forecast accuracy, AGRICAF usually achieves high accuracy in its price forecasts, especially for short forecast horizons, coloured in dark purple. Another factor that can impair the model's performance is high market instability, as exemplified by the periods of extreme price changes, and show lower overall performance. Such events characterised by abrupt or extremely high positive or negative price changes can result in large differences between the observed price (represented by the black line) and the forecasted price (depicted by the purple lines). 

\subsection{Understanding Agricultural Commodity Price Fluctuations: Exploring the Factors Behind AGRICAF's Forecasting Results}
In this article, we begin with a broad explanation and then zoom in to get a more detailed, specific understanding of the forecasts generated by the model.

After obtaining the forecasted prices, we inspect the factors which have driven these results.
For the purpose of this article, we start by the most general explanation and, gradually, zoom in to gain detailed, more specific, explanation of different impacts which, together, led to the forecasts obtained by the model.
\subsubsection*{General Interpretation}
\begin{figure}
    \centering
    \begin{subfigure}[b]{0.49\textwidth}
        \includegraphics[width=\textwidth]{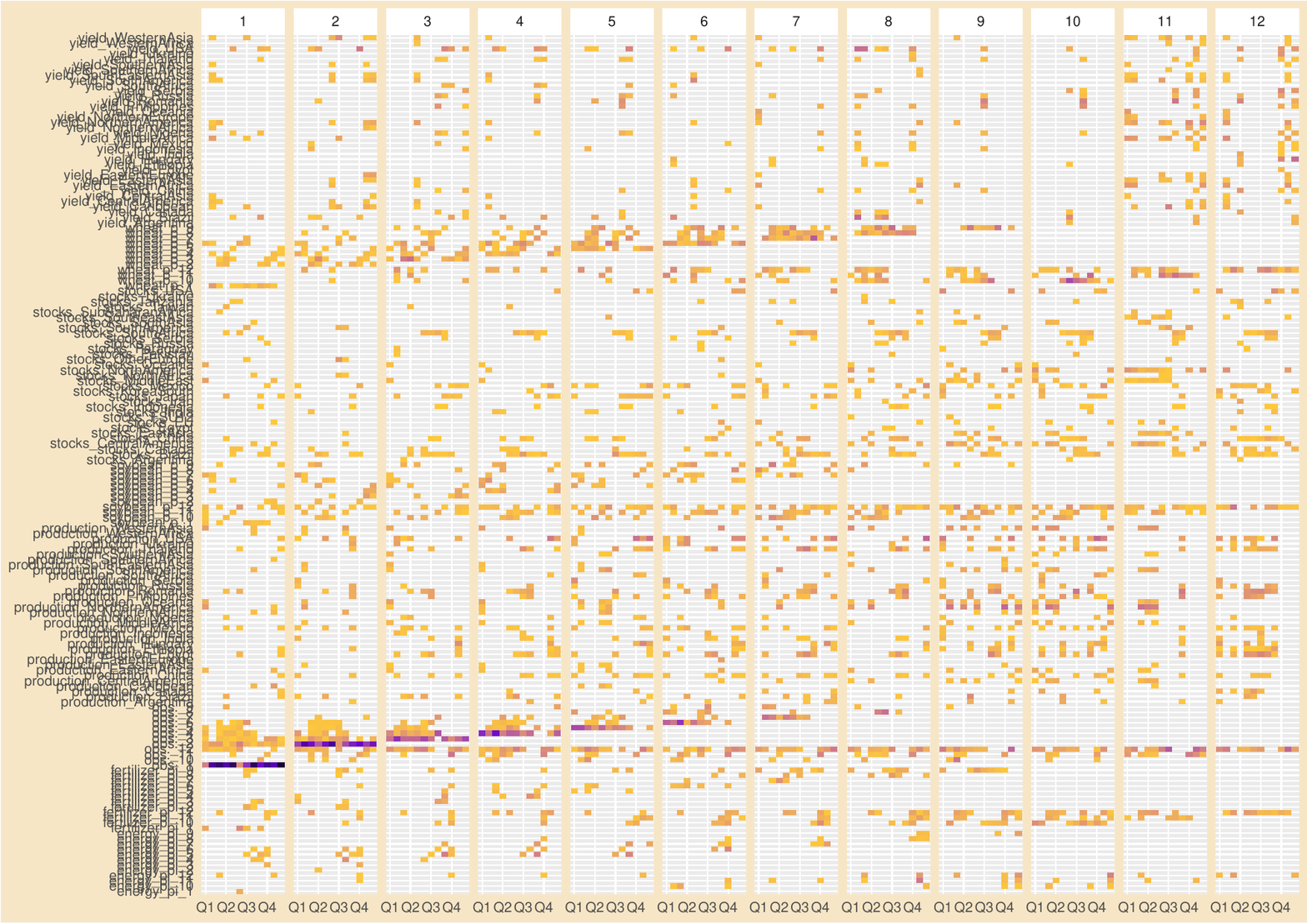}
        \caption{Maize}
        \label{fig:maize}
    \end{subfigure}
    \hfill
    \begin{subfigure}[b]{0.49\textwidth}
        \includegraphics[width=\textwidth]{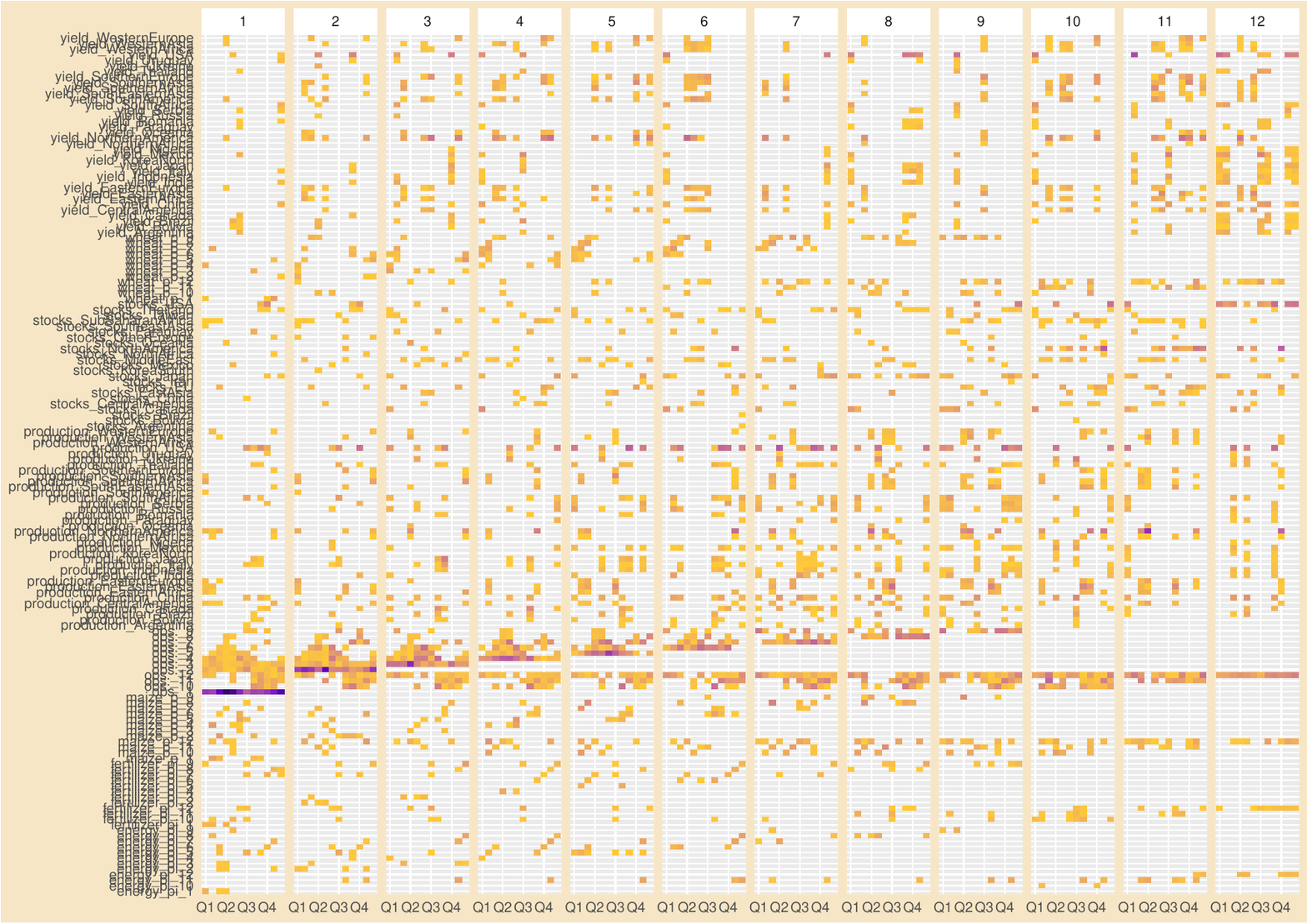}
        \caption{Soybean}
        \label{fig:soybean}
    \end{subfigure}
    
    \begin{subfigure}[b]{0.49\textwidth}
        \includegraphics[width=\textwidth]{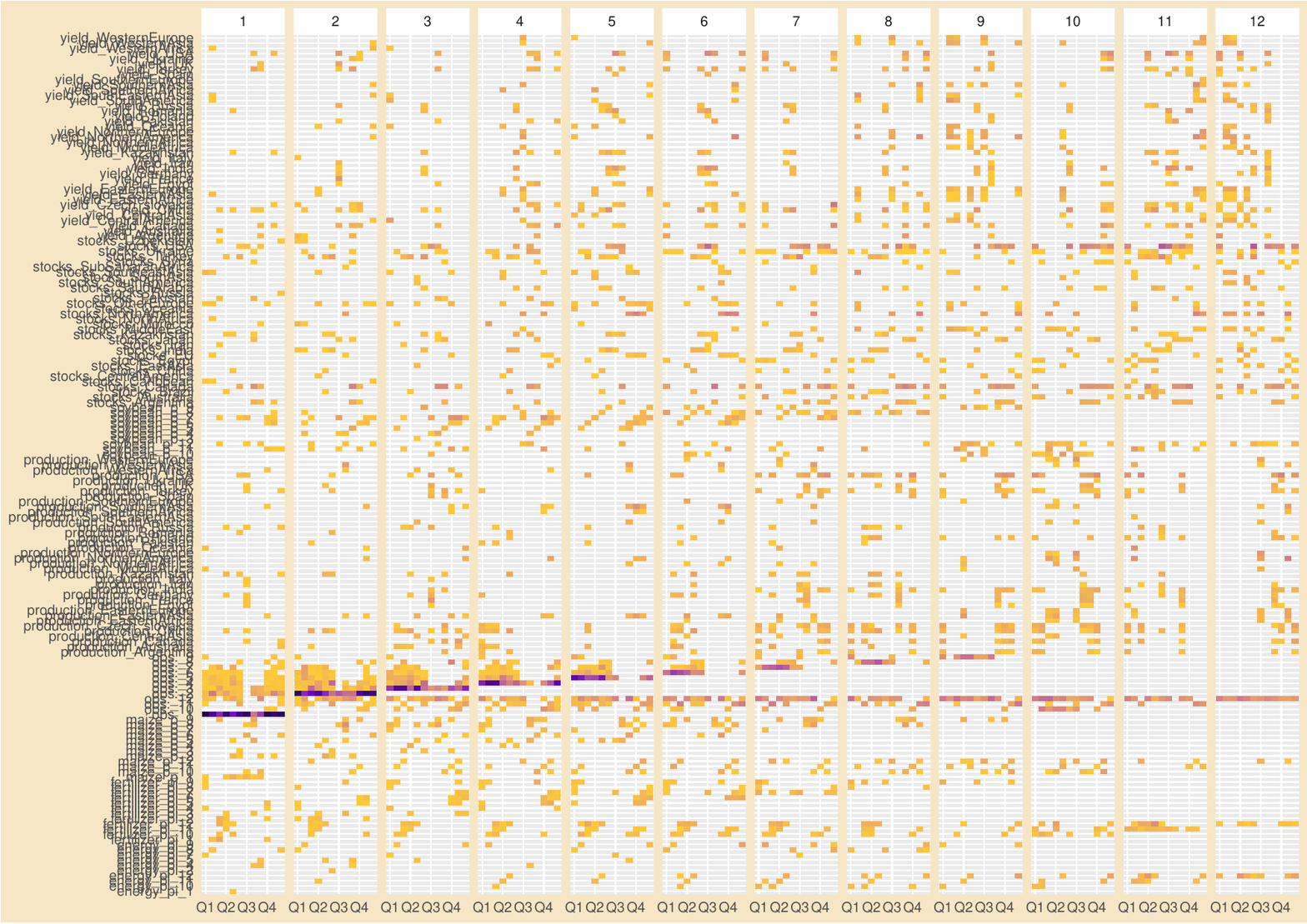}
        \caption{Wheat}
        \label{fig:wheat}
    \end{subfigure}
        \hfill
    \begin{subfigure}[b]{0.49\textwidth}
        \includegraphics[width=\textwidth]{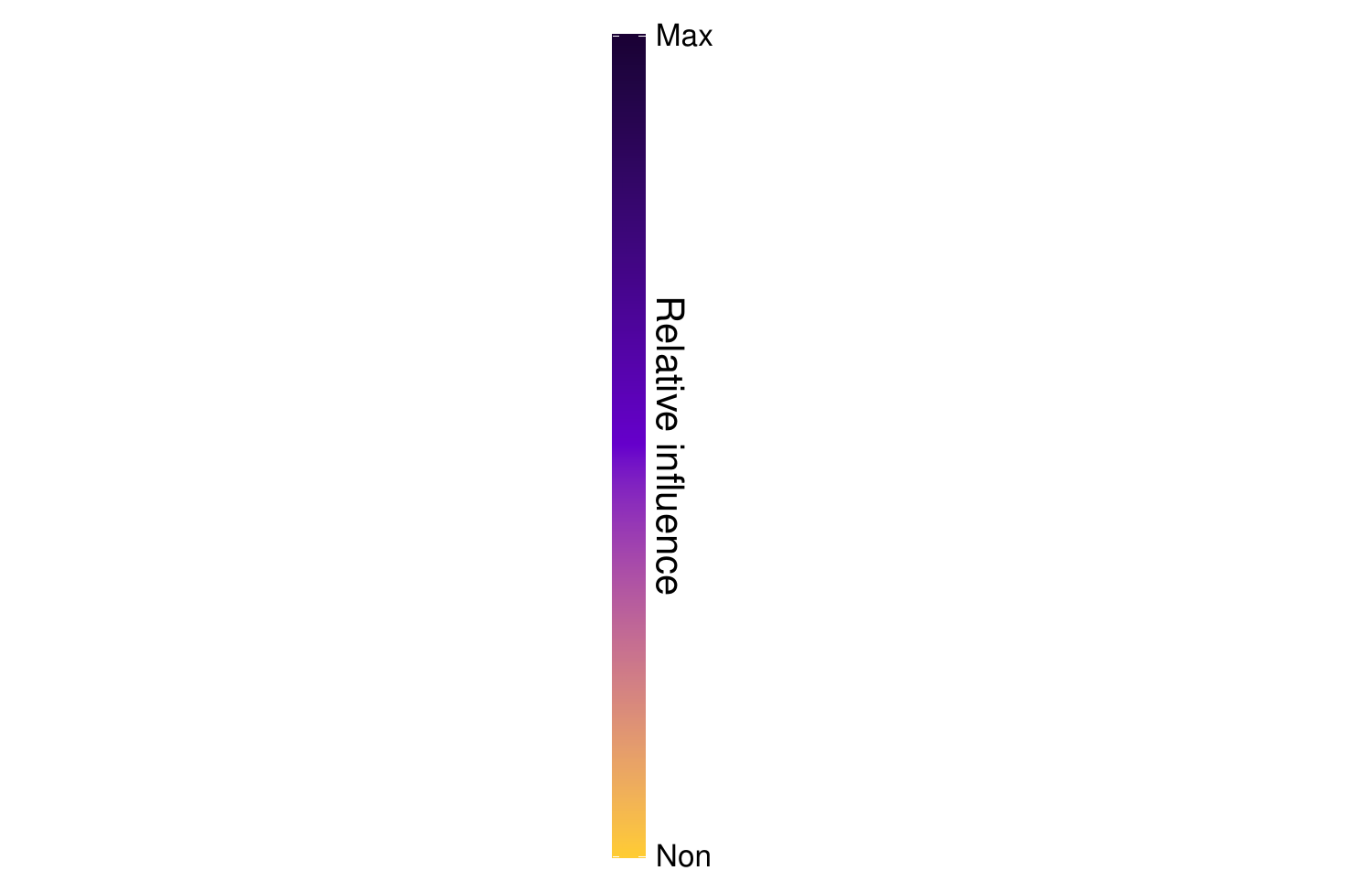}
        \label{fig:legend}
    \end{subfigure}
    \hfill
    \caption{Assessment matrix showing the median relative influence (Shapley value) of annual changes in selected features (vertical axis) for monthly price forecasting of maize (a), soybean (b), and wheat (c). The months are marked on the horizontal axis as quarters (Q1, Q2, Q3, Q4), with panels representing 1-12 month forecasting horizons. Dark purple indicates factors with high marginal influence, while orange represents factors with a low marginal influence.}
    \label{global_matrix}
\end{figure}

The broadest and least detailed level of explanation is the general overview of forecasting results. This interpretation provides an all-year perspective on the marginal effects of simultaneous changes in key factors on a commodity's monthly price changes across 12 forecast horizons. Presented as three matrices, Fig.~\ref{global_matrix} illustrates a global, agnostic view of the forecasted prices for the agricultural commodities in focus.

Divided into 12 panels, the plot's vertical axes lists of all the features retained at the final dataset, after rigorous screening (see \hyperlink{methods1_cmaaf_stages}{Stages 1 and 2} in the Methods section). Each panel represents one of 12 forecasting horizons for monthly price forecasts, with each forecast using up to 19 explanatory variables.\footnote{
A complete and detailed list of variables and indices is given in Tab.~\ref{tab_var_list}
} 
These variables represent the most impactful factors, based on median of the absolute Shapley values for a given feature, which provides a measure of the marginal impact of that feature on the forecasted commodity price, as forecasted by the best-performing model relative to a month's price change and forecast horizon. This value helps in understanding how the typical (median) impact of a specific feature compares to the overall contribution of all features. A higher ratio indicates that the feature has a more significant typical impact relative to the overall feature contributions.
The horizontal axes correspond to the quarters of the calendar year, with each representing a different month. 
The coloured tabs indicate a single feature's impact to each individual price change prediction. Impact level ranges from 0 (bright yellow), indicating that including the feature does not impact the model's prediction performance, to a maximum of 1 (dark purple), indicating that out of all features, only one had a positive impact on the model's forecasting performance. Gray tabs belong to features which have found to be price drivers, but in other period of the year or a different forecast horizon. It should be noted that negative contributions are possible but very rare in our forecasting tool, thanks to the rigorous variable screening performed in the early stages of AGRICAF.

From an overview of the depicted data, several key insights emerge. Firstly, across all crops, there are seasons where similar features drive price changes, as indicated by rows with connected coloured tabs. Secondly, the majority of the selected features are coloured yellow or bright orange, indicating a relatively low impact on forecasted prices across all crops. This highlights the influence of using a diverse range of factors to achieve a reliable forecasting model capable of accurately predicting specific crop price fluctuations. Notably, in shorter time horizons, there are fewer features with substantial impact on prices, with few variables showing high relative influence, coloured in purple. A closer look reveals that these variables are financial, such as historical prices of the same or other commodities. As the forecast horizon increases, the relative influence of these individual financial variables diminishes, and other factors related to agricultural supply become more prominent. This shift suggests that a broader array of factors impacts price changes as the forecast horizon extends, emphasising the complexity and dynamic nature of agricultural commodity pricing. 

\subsubsection*{Local Interpretation}
Requesting a specific month to forecast with AGRICAF allows for a deeper understanding of the model's results and the learning process behind them. In the global trade of agricultural commodities, September marks the beginning of the trade year for maize and soybean in the USA, their largest producer. Similarly, July starts the trade year for wheat in most producing countries, including major producers like China, Russia, and Ukraine. These months significantly influence the supply and demand dynamics in their respective markets \citep{cme2024education,oecd_fao2022}. For each of these three commodities, we present an analysis of the significance of the chosen features on the accuracy of our price forecasts, performed 12 months in advance (refer to \hyperlink{03_methods}{Stage 3} in the Methods section).

\begin{figure}
    \centering
    \begin{subfigure}[b]{0.49\textwidth}
        \includegraphics[width=\textwidth]{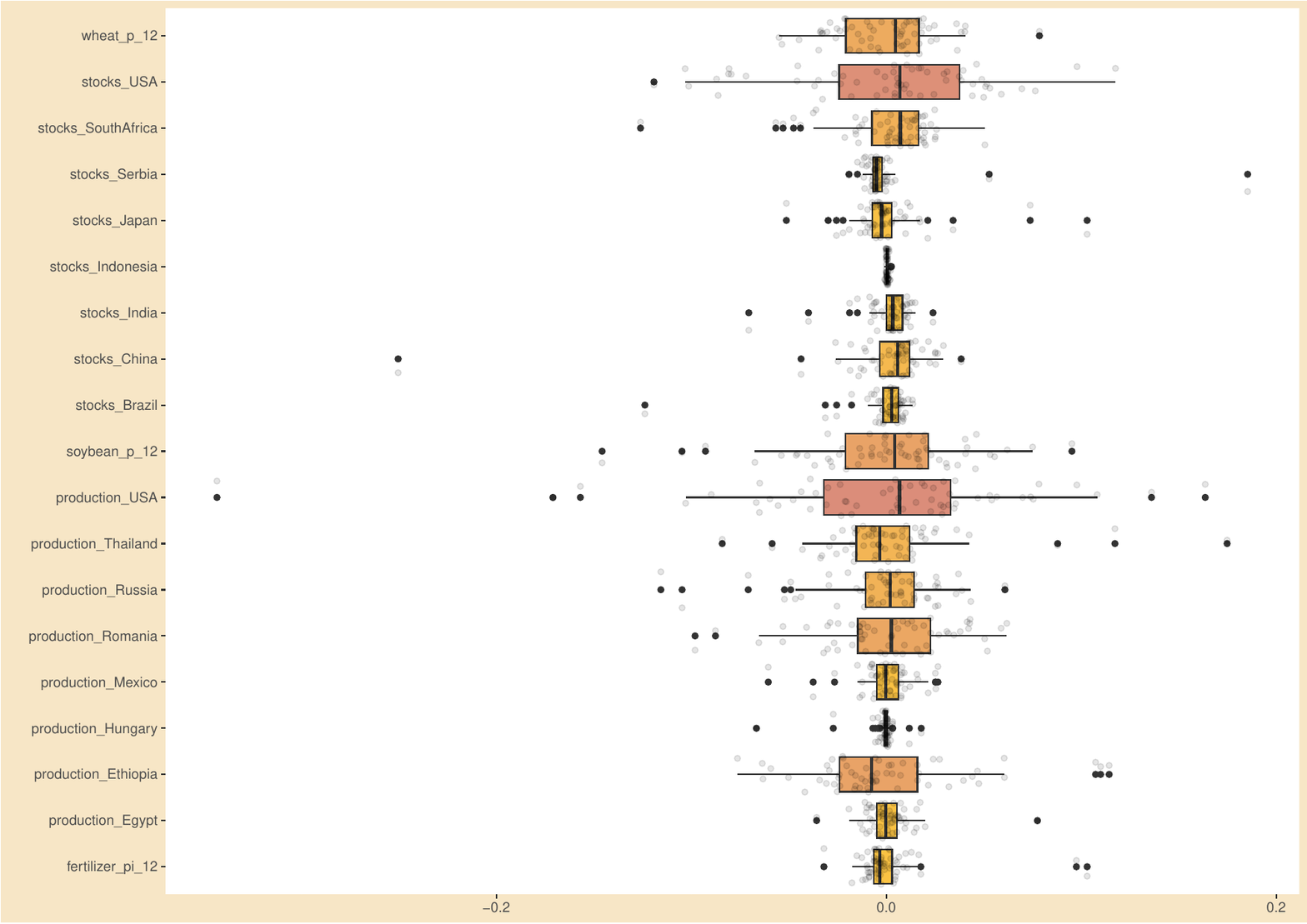}
        \caption{Maize}
        \label{fig:maize}
    \end{subfigure}
    \hfill
    \begin{subfigure}[b]{0.49\textwidth}
        \includegraphics[width=\textwidth]{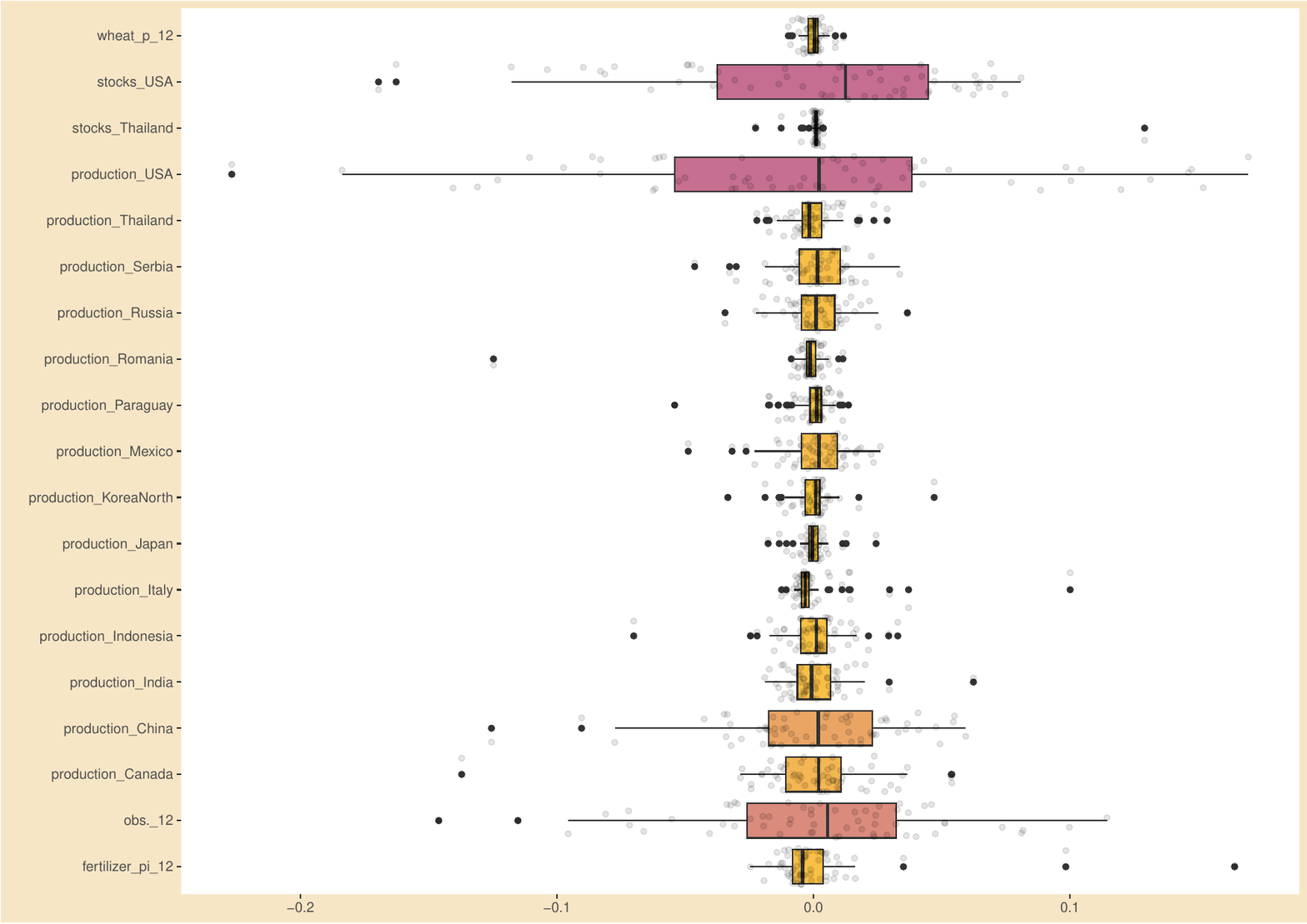}
        \caption{Soybean}
        \label{fig:soybean}
    \end{subfigure}
    
    \begin{subfigure}[b]{0.49\textwidth}
        \includegraphics[width=\textwidth]{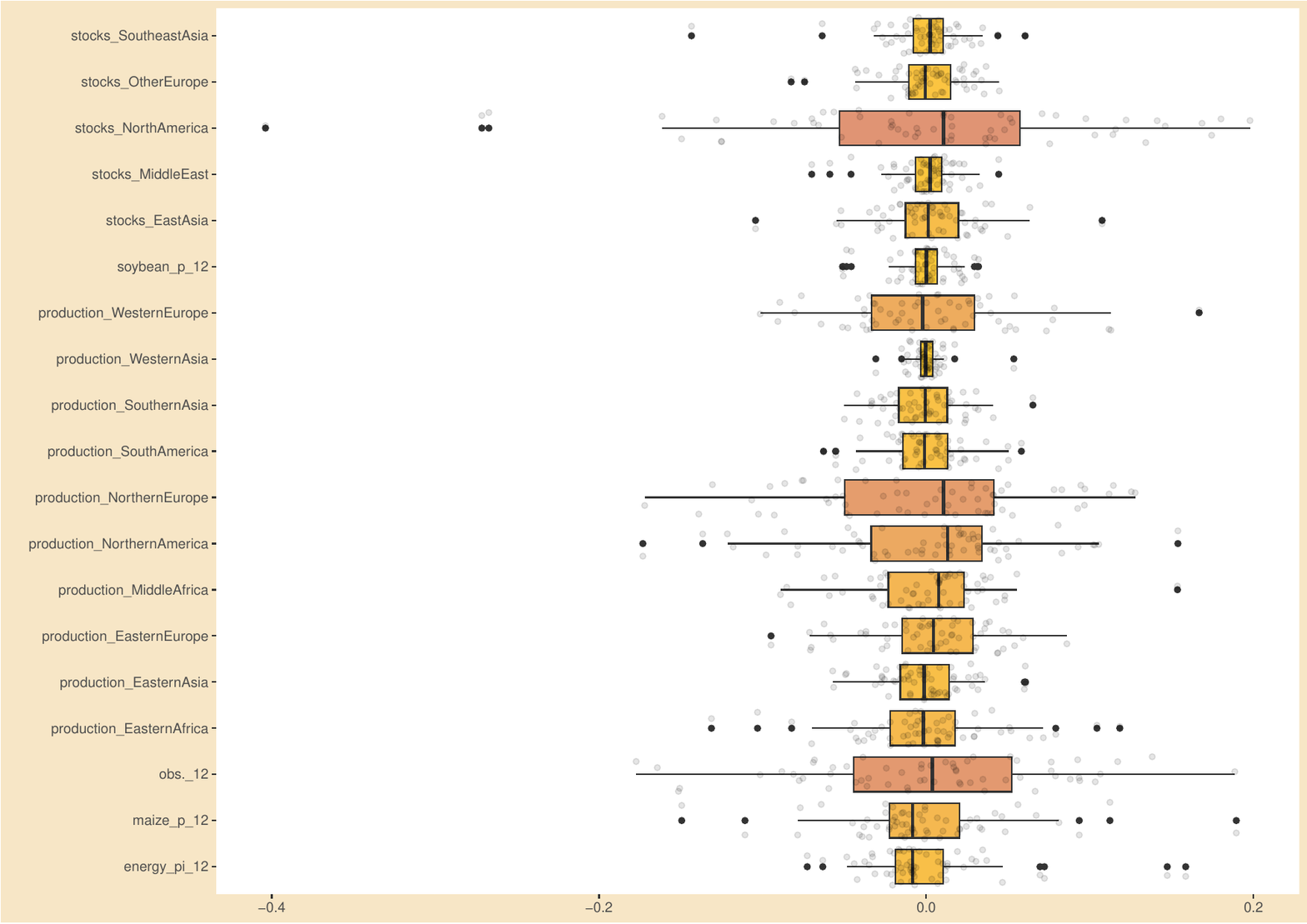}
        \caption{Wheat}
        \label{fig:wheat}
    \end{subfigure}
    \hfill
    \begin{subfigure}[b]{0.49\textwidth}
        \includegraphics[width=\textwidth]{graphics/influence_legend.pdf}
        \label{fig:legend}
    \end{subfigure}
    \hfill    
    \caption{Median relative influence (Shapley value) of annual changes in selected factors (vertical axis) for monthly price forecasting of maize (a) and soybean (b) in September, and wheat (c) in July. The box-plots display the distribution of Shapley values, revealing the impact of each factor selected by AGRICAF on the forecasted relative price change. Grey points represent individual Shapley values attributed to specific predictor variables.}
    \label{local_varImp}
\end{figure}

Fig.~\ref{local_varImp} shows the contribution of the selected variables, which are coloured based on their level of marginal influence, i.e., the contribution of individual factors on the predicted price variation.
At the top are the September maize (left) and soybean prices, which are characterised globally by a rather high concentration in terms of relative influence. Fig.~\ref{local_varImp} reveals that changes both USA's supply factors - production and stocks - exert the strongest influence on price change predictions. In the case of maize, these factor are valued together 27\% (14\% and 13\%) of the total influence. 
As of for soybean's price in September, USA is ranked very high with a total of 42\% influence (21\% for each, production and stocks), making a mean value of 0.06 and 0.05 relative influence.
For both commodities, when forecasting 1 year ahead, USA is ranked as at least one of the two most impactful factors affect the global price changes for all months. 
Similar to maize and soybean, the USA has the highest marginal influence seems to have high influence on the July wheat prices during the start of the local trade year; this time through stocks (13\%) and production changes in Northern America (8.5\%). It is closely followed by other impactors: historic prices (12\%) and Northern Europe's production (11\%).

Close analysis of other months and forecast horizons reveals that the influence of various features on the forecasted price of all three commodities varies across months and forecasting horizons. The most impactful features vary significantly between short-term (1 to several months ahead) and longer-term forecasts. The impact is also notably different from month to month within the trade year. Key observations include the dominance of certain features like historic prices in short-term.  
As the forecasting horizon extends, the relative importance of individual features fluctuates, reflecting the increasing complexity of long-term price predictions.

\begin{figure}
    \centering
    \begin{subfigure}[b]{0.49\textwidth}
        \includegraphics[width=\textwidth]{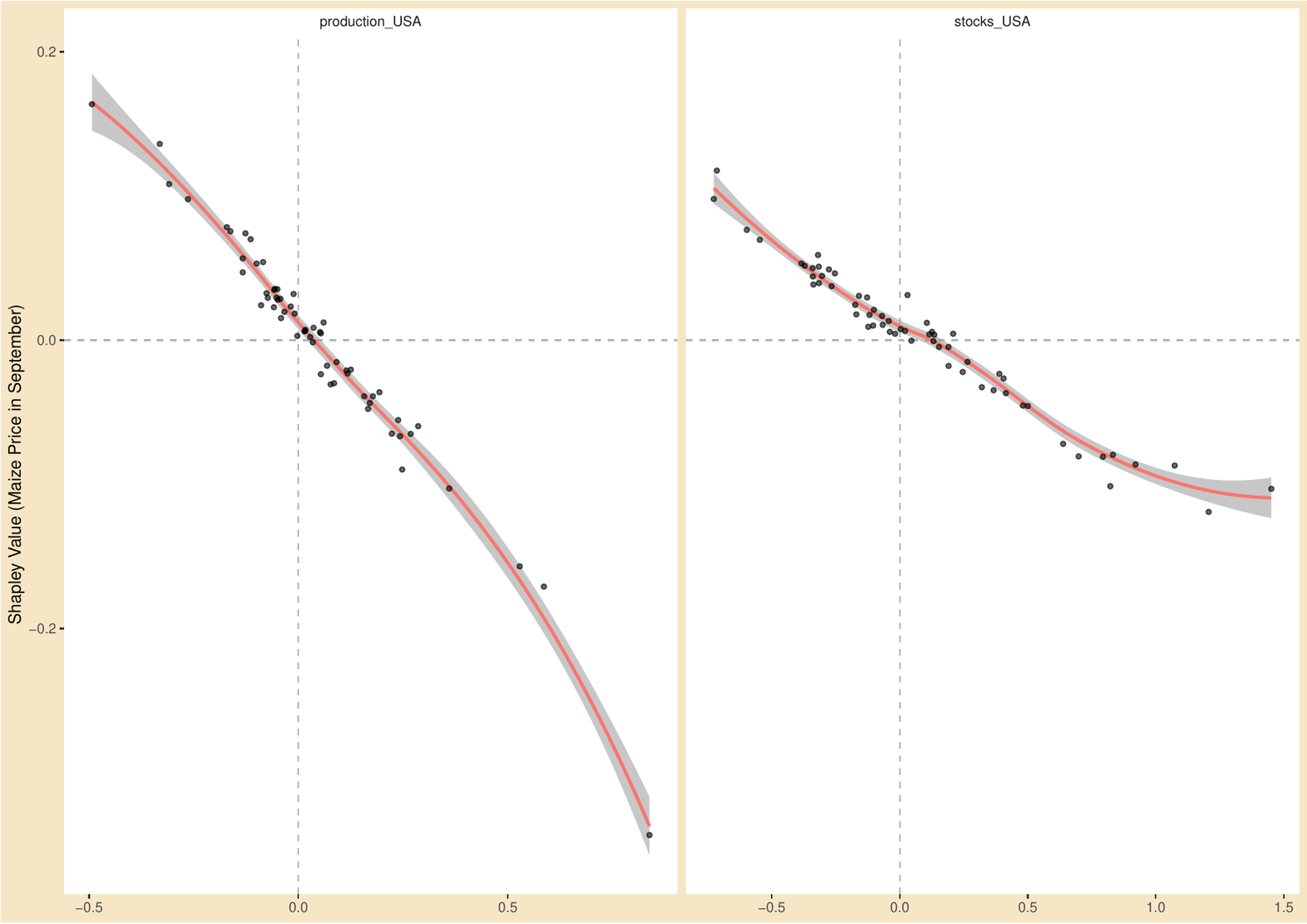}
        \caption{Maize}
        \label{fig:maize}
    \end{subfigure}
    \hfill
    \begin{subfigure}[b]{0.49\textwidth}
        \includegraphics[width=\textwidth]{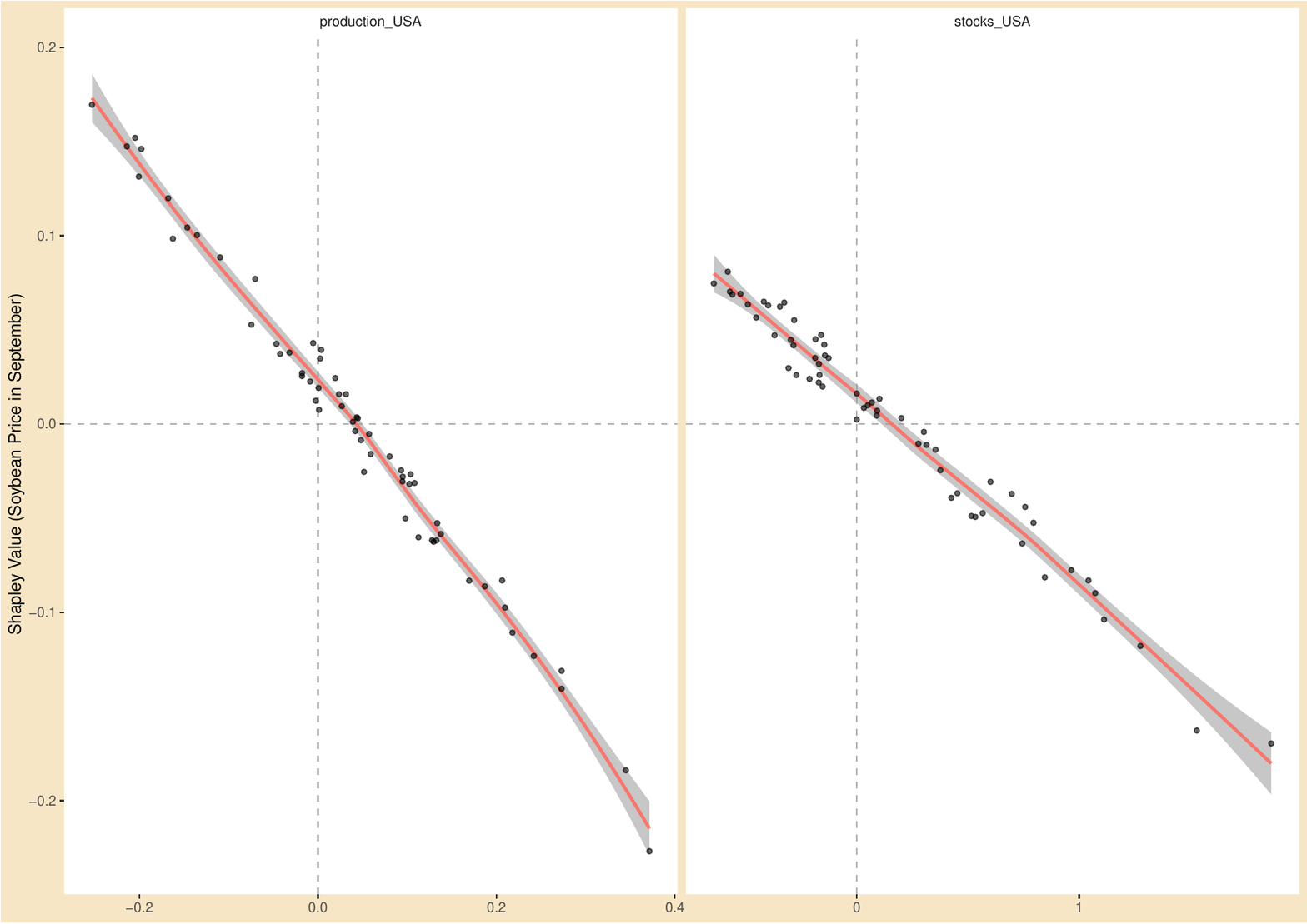}
        \caption{Soybean}
        \label{fig:soybean}
    \end{subfigure}
    
    \begin{subfigure}[b]{0.49\textwidth}
        \includegraphics[width=\textwidth]{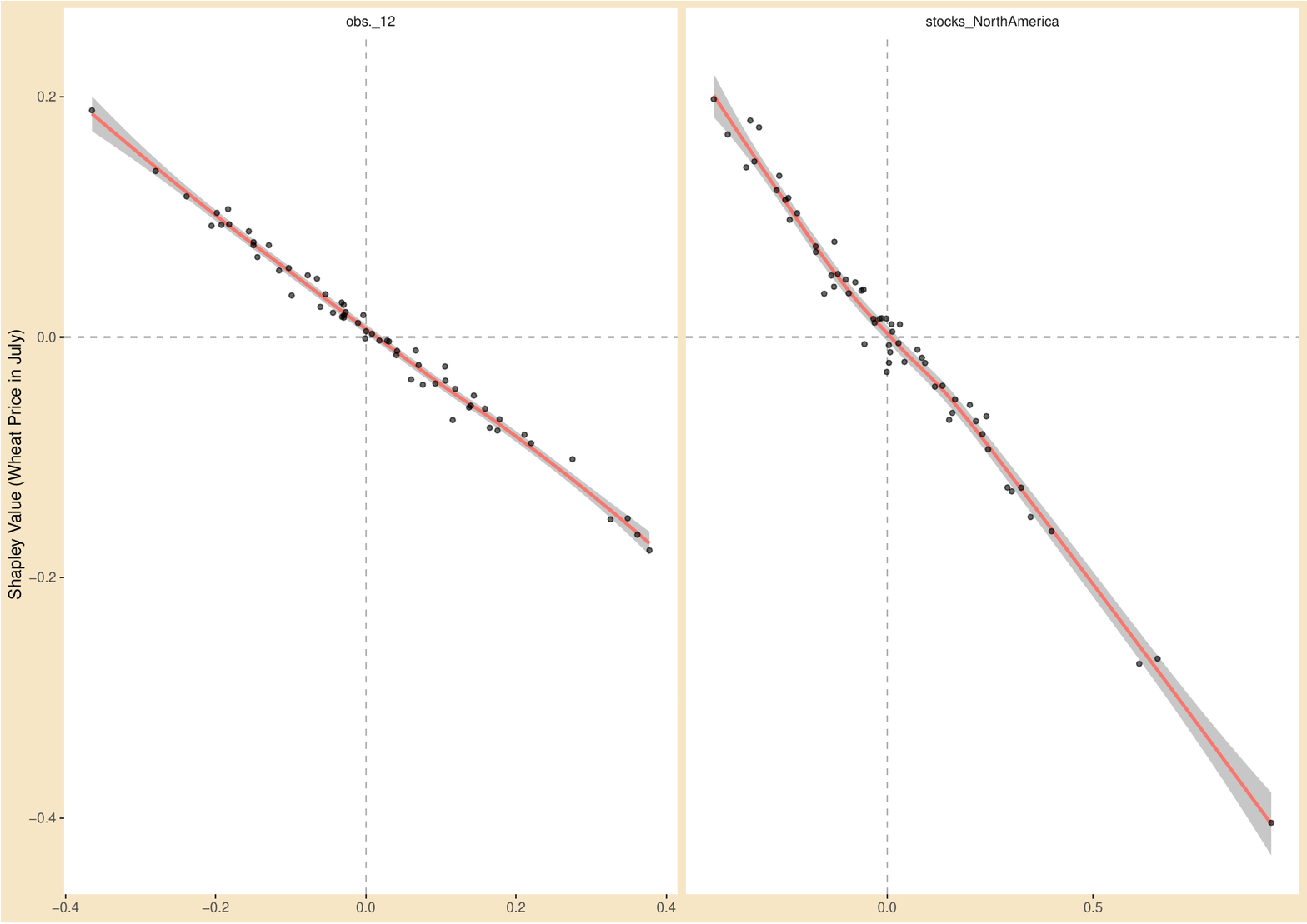}
        \caption{Wheat}
        \label{fig:wheat}
    \end{subfigure}
    \caption{Shapley-based partial dependence plot (PDP) for maize (a), soybean (b) and wheat (c). Individual feature values are denoted by black points scattered along the x-axis, with their corresponding Shapley values illustrated on the y-axis. A red smoothing curve represents the PDP, illustrating the relationship between relative changes in the most influential factors and the projected relative AC price changes for the selected month. Each PDP is based on the forecasting model with the highest accuracy. The grey areas indicate the 95\% confidence intervals.}
    \label{fig:local_pdp}
\end{figure}

Exploring higher resolution of model agnostic, Fig.~\ref{fig:local_pdp} highlights the two predictors with the greatest relative importance. 
Changes in USA's production and stocks have the highest influence in September's maize (a) and soybean's price (b). For wheat's July price the factor with the highest influence the derived from Northern American stocks level, and is closely followed by wheat's own historic prices and production in Northern Europe (c).

The three segments of the partial dependence plots (PDP) exhibit inverse relationships between the forecasted price and the feature, meaning that an increase (decrease) in any of these factors tends to result in a decrease (increase) in the corresponding price change. 
In general, the USA seems to have strong impact on the three commodities at the local new-crop (beginning of the local trade year) time, despite the fact that it is not the biggest producer nor the biggest consumer of wheat. 
For the latter, market powers are also very prominent, with the PDP highlights the influence of the previous year's price.
Another observation is that the PDP red line intersects the X-axis at the point where there is no change in USA production, relative to both maize and soybean prices. 
Comparatively, maize and soybean prices are more sensitive to production changes, while wheat prices are significantly affected by historical price trends and stocks levels.

Finally, we detail a specific instance where our model sheds light on market shifts that influenced price changes during particular events. We concentrate on the wheat price fluctuations in July 2022, which occurred at the beginning of the start of the Russia-Ukraine conflict, a period when wheat prices reached their peak. Our aim is to assess the effectiveness of AGRICAF on a case study that emerged unexpectedly and was not included in the initial training data.

\begin{figure}
        \centering
    \begin{subfigure}[b]{0.49\textwidth}
        \includegraphics[width=\textwidth]{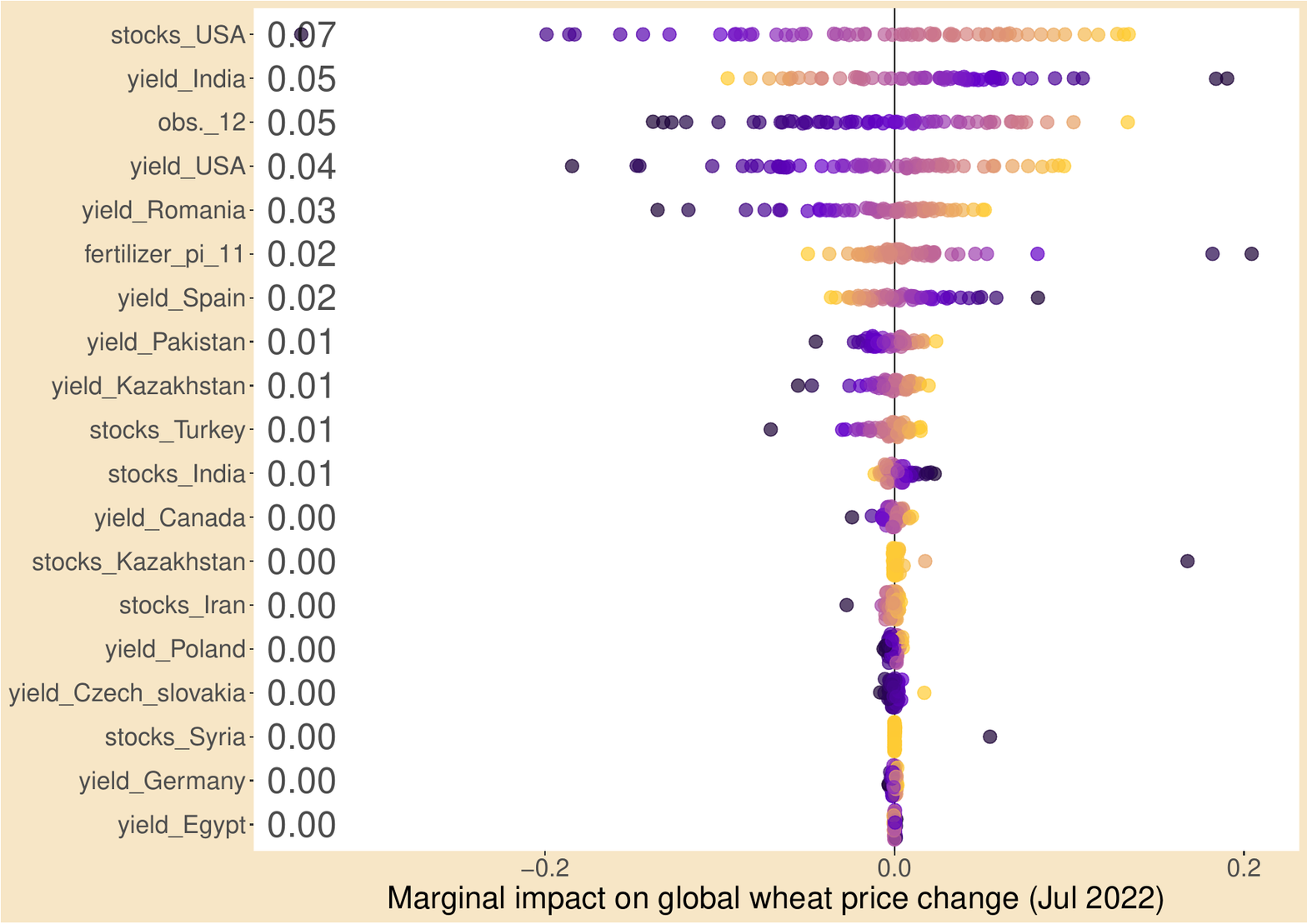}
        \caption{Forecast horizon = 11 months}
    \end{subfigure}
    \hfill
    \begin{subfigure}[b]{0.49\textwidth}
        \includegraphics[width=\textwidth]{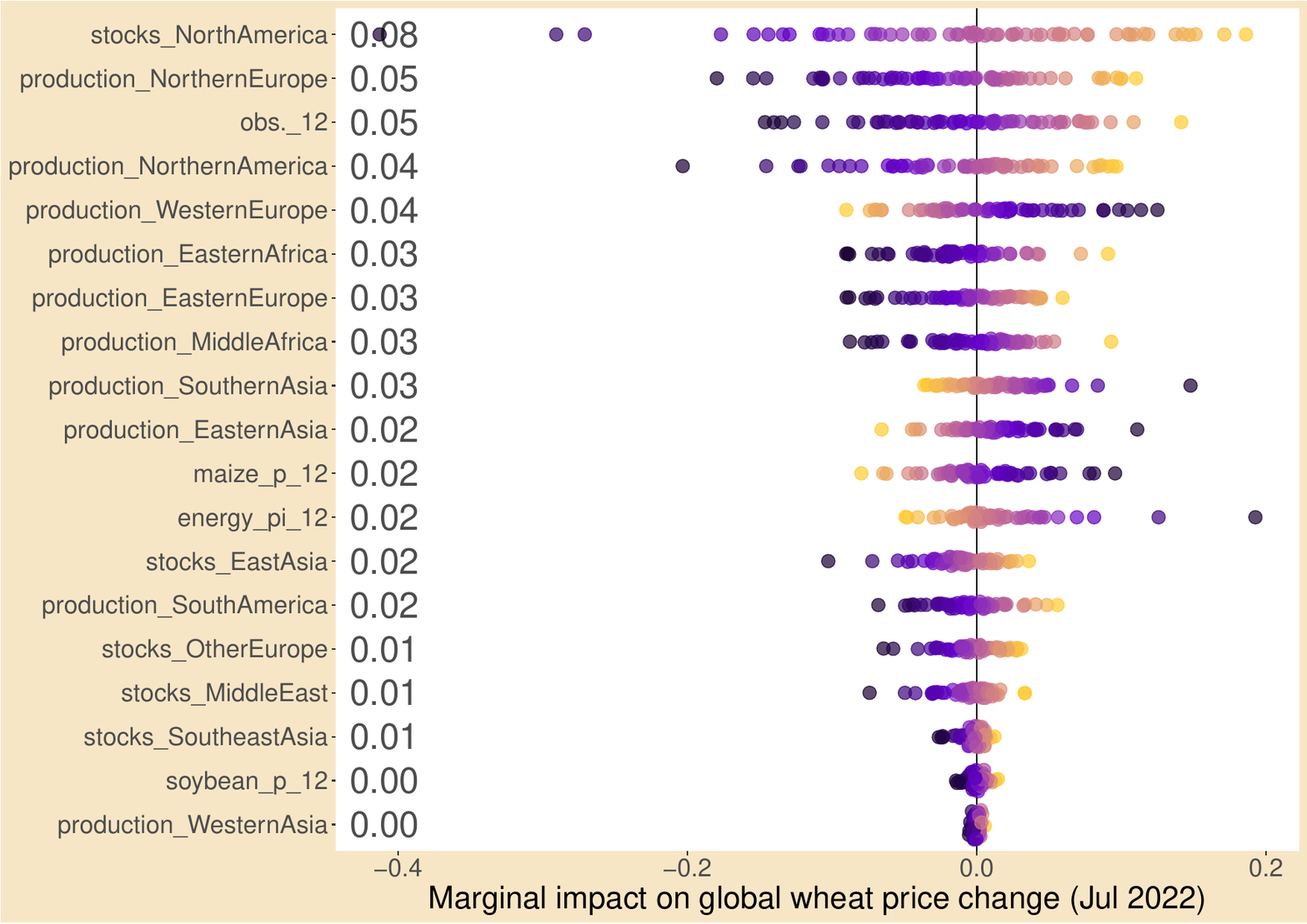}
        \caption{Forecast horizon = 12 months}
    \end{subfigure}
       \hfill
    \begin{subfigure}[b]{\textwidth}
        \includegraphics[width=\textwidth]{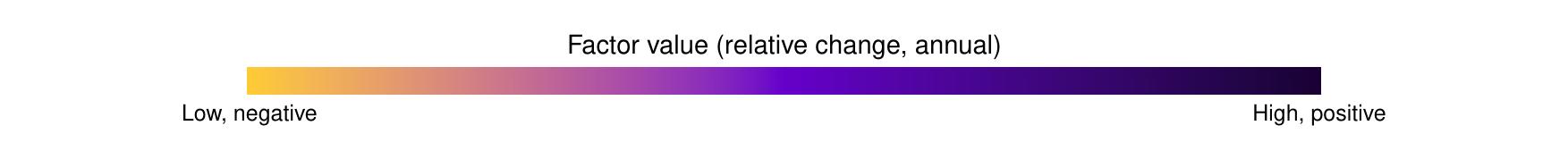}
    \end{subfigure}

    \caption{Marginal impact on global wheat price change in July 2022, as forecasted in two horizons: 11 and 12 months ahead. Ranked from top to bottom, the graph represents the influence of highest ranked factors on forecasts of wheat price changes. Each point, coloured from deep purple (high value) to orange (low value), indicates the impact level of changes in a factor on price changes. Points are aligned on the X-axis based on their Shapley values, with extreme negative impacts to the right and vice versa. The numbers on the Y-axis summarise the average impacts of the factors.}
    \label{case_study}
\end{figure}

The AGRICAF application, shown in Fig. ~\ref{case_study}, selected the XGBoost model with a linear booster, which resulted in the smallest Mean Absolute Error (MAE) and the highest Relative Advantage (RA). 
The actual relative price change in that month was 21.7\% higher, compared to July 2021. The forecast made one year in advance predicted a 21.8\% increase, while the forecast made eleven months in advance predicted a 19.8\% increase.
In the 12-month forecast, the factors with the highest influence were stock levels in Northern America, historical wheat prices from twelve months prior, and various production metrics across different regions, indicating a significant reliance on supply data and past prices. The 11-month forecast showed a similar pattern, with stock levels in the USA and historical prices again being prominent factors. along with yield data from key wheat-producing countries like India, the USA, and Romania. These influences reflect the substantial impact of stock levels, historical trends, and production capabilities on wheat price forecasts.

\section{Discussion and Conclusions}
This study explores the possibility to accurately forecast price changes of internationally traded agricultural commodities (ACs), using only publicly available data and a combination of econometric and explainable machine learning (XML) methods. The AGRICAF (Agricultural Commodity Analysis and Forecasts) methodology demonstrates its applicability across three key commodities: maize, soybean, and wheat. 
These prices, updated monthly since January 1960 at the \cite{worldbankprices}, are explained by other publicly available data which are also reported and regularly updated by reliable sources. Together, they form the foundation of our analysis.

Machine learning models are often praised for their forecasting accuracy in stable markets, where historical data patterns persist \citep{athanasopoulos2024forecast, makridakis2018Statistical}. However, agricultural commodities present a unique challenge due to their natural volatility and susceptibility to external influences such as weather, geopolitics, and supply-demand shifts \citep{headey2010reflections}. When these conditions fluctuate, the reliability of machine learning models can decrease, especially over extended time frames. This study highlights that while XML methods can effectively predict short-term price changes, their accuracy diminishes as the forecast horizon lengthens.

AGRICAF advances AC price forecasting by providing accurate predictions within a 1-12 month horizon using only publicly available data. It forgoes the usually applied expensive, proprietary datasets \citep{hernandez2010examining, gouel2012pricestability} and instead leverages innovative cross-validation techniques to capture complex market dynamics. The methodology combines various statistical tests, analytical approaches and the application of XML and time series models, ensuring robustness and interpretability. By making price forecasts accessible to a broader audience—regardless of budget, language, or expertise—AGRICAF fills a critical gap in AC market analysis.


In our study, AGRICAF  provided accurate predictions for the upcoming year, with its performance most reliable for shorter time frames. As forecast horizons extended, accuracy generally declined, particularly when predicting prices 12 months ahead. This suggests that AGRICAF is most effective for near- and medium-term forecasts, and users should be cautious when relying on it for longer-term predictions. Extreme market events, such as abrupt price changes, further challenged the accuracy of our model, underlining the inherent difficulties in forecasting agricultural prices despite sophisticated methodologies.
Notably, the largest forecast errors for each commodity were associated with significant market disruptions such as extreme weather events, geopolitical tensions, or global supply chain disruptions. For instance, major events such as the 2010 Russian drought \citep{zampieri2016global,wegren2011food}, the 2021 COVID-19-related supply chain disruptions \citep{espitia2020covid,cariappa2022covid}, and the Russia-Ukraine war \citep{glauber2023russia} contributed to substantial forecasting errors at the year-ahead horizons.

The inclusion of different types of models and inputs showed the benefits of careful data analysis and model adjustment, which often yielded forecasts more accurate than those achieved using advanced time series models (ARIMA, TBATS, VAR). These results highlight the importance of continuous refinement and adaptation of forecasting techniques to master the complexities of agricultural commodity markets.

We also examined the drivers behind these forecasts. Understanding these underlying factors is vital, especially for non-specialists. These insights into the model’s mechanism and the specific market dynamics allow users, such as policymakers and farmers, to make informed decisions based on the forecasts, rather than blindly following them. This transparency can improve strategies around food security and social equity, allowing stakeholders to anticipate and adapt to market changes more effectively.

Our analysis of feature influence (and importance) revealed complex patterns and relationships among features and forecasted prices of ACs, especially when forecast horizons become longer. They have showed that similar factors often influence price changes across different seasons for all crops. However, most features demonstrated low impact. In the maize and wheat markets, there was an apparent strong impact of very few features, especially in short horizons, necessitating further investigation into the identification of these market forces.

Going into higher resolution, we showed a comparative reflection of the relative influence of each variable on the predicted price; and descriptive example, where we outlined the average behaviour of a chosen variable throughout the time series \citep{lundberg2017shap}. 
Our analysis revealed how the nature of each commodity significantly influences the type, quantity, and impact level of the factors affecting its price over time, leading to considerable variation in the initial and finally-selected factors between commodities and throughout the market year, relative to the forecast horizon. 

When examining the influence of factors across months, no consistent pattern emerges that would suggest a specific month significantly alters the relative influence's distribution for either crop. However, certain seasons show superior forecasting accuracy, particularly for longer forecast horizons. This indicates that the relative influence of forecasting features changes throughout the year, depending on the month in question and the period-length between the recorded change in the feature and the desired period to forecast. However, individual features such as fertiliser prices and historical crop prices (observed prices) exhibit higher degrees of influence for short horizons, while supply related variables have higher impact in the longer-run.

For forecasts with a one-year horizon, distinct patterns of feature influence emerge across different crops. 
In the case of maize, production-related factors from key producing regions such as the USA, Egypt, and Central-Eastern European countries play a significant role in price forecasts. Notably, the USA's influence is particularly strong during the first half of the local trade year, starting in September. As the year progresses, the influence of Egypt and European countries becomes more pronounced. Additionally, historical prices—both for maize itself and related inputs like fertilizers and soybeans (another biofuel crop)—also have a considerable impact on forecasts throughout the year.

In contrast, soybean price forecasts tend to be influenced by fewer factors at any given month. From September to May, the dominant drivers of price forecasts are USA supply levels, i.e., production and stocks levels of the biggest exporter of soybean. Interestingly, production levels in Mexico and Eastern Asian countries—world top importers of soybean—become more influential during the remainder of the year. The influence of historical prices is also present throughout the entire year.

Wheat forecasts, on the other hand, show a consistent reliance on historical prices, highlighting the importance of financial and past market trends in determining future prices. In addition to historical data, USA stock levels and production in Eastern European countries also significantly impact wheat price forecasts. These qualitative differences across maize, soybean, and wheat put light on the varying market structures and dependencies that shape price dynamics in each sector.

A case study tested the actual applicability of AGRICAF to the real world, during the occurrence of an unexceptional event. We applied two forecast horizons (11 and 12 months) and considered the wheat prices in July 2022. 
We deliberately chose an extreme period, in which the market experienced great disruption from different directions: The COVID-19 pandemic caused significant disruptions in trade and supply chains \citep{cariappa2022covid}. Geopolitical tensions, notably the war in Ukraine, have also had a substantial impact. Ukraine and Russia are major wheat producers, and the conflict disrupted their wheat exports, leading to global supply shortages and increased prices \citep{falkendal2021grain,glauber2023russia}. Moreover, major wheat-importing regions, faced heightened demand for wheat due to their reliance on imports, especially from Ukraine \citep{devadoss2024impacts}. This surge in demand, worsened by the preceding global stock shortages, contributed significantly to the observed increase in wheat prices. Additionally, climatic events such as droughts in major wheat-producing regions have further strained supply levels. These results underscore AGRICAF's capability to provide actionable forecasts even during periods of extreme market volatility.

For low-income countries, the insights from AGRICAF enable targeted strategies to support smallholder farmers, who are disproportionately affected by price volatility, by providing advanced market information and helping to ensure fairer pricing structures. Ultimately, AGRICAF’s accessible forecasts can foster a more resilient food system, contributing to food security and promoting social equity.

AGRICAF offers a path toward more efficient and socially responsible agricultural forecasting, contributing to a fairer and more sustainable global food system. Future work could enhance the methodology by incorporating additional data types, such as policy or geospatial information, and expanding its scope to include more commodities and region-specific forecasts. To maximise accessibility, AGRICAF may develop into an app or online platform, similar to tools like the \href{https://fpma.fao.org/giews/fpmat4/#/dashboard/tool/international}{FPMA} or \href{https://www.foodsecurityportal.org/}{FSP}. Despite its limitations, AGRICAF is a powerful tool for medium-term AC price forecasting, much like weather forecasting in the 20th century, enabling informed decision-making in global food markets.

Ultimately, AGRICAF’s accessible, transparent, and adaptable approach not only enhances food market management but also contributes to a more equitable and inclusive global food system. By providing data-driven insights accessible to a range of stakeholders, AGRICAF directly contributes to the global Sustainable Development Goals (SDGs) aimed at ending hunger, improving food security, and more equitable society. In this way, AGRICAF has the potential to make meaningful role in achieving a fairer, more resilient global food system, where stakeholders at every level are equipped to make informed, timely, and impactful decisions.

\section{Acknowledgements}
This work was supported by the European Union’s Horizon research and innovation programme under the Marie Skłodowska-Curie Actions fellowship (grant agreement No. 101111405 – project A comprehensive method for medium-term analysis and forecasting (CMAF) of global monthly prices of agricultural commodities).

I sincerely thank Anne Goujon, Sören Lindner, Sebastian Poledna, and Serguei Kaniovski for their valuable feedback. I am also grateful to Christiane Pohn-Hufnagl for her unwavering support beyond research and for her invaluable guidance. I extend my deepest appreciation to my PhD advisor, David Makowski, for his continuous support and insightful feedback. Special thanks to Jesus Crespo Cuaresma for his generous advice and warm welcome into the WU community, and to Clelia Minaudo for her dedicated support as my project adviser.

\newpage
\bibliographystyle{chicago}
\bibliography{bibliography}

\makeatletter
\renewcommand{\@biblabel}[1]{\quad#1.}

\label{appendix}
\begin{subappendices}
\newpage
\section[Appendix A]{Appendix: Data \& Variables}\label{appendix_data}
\begin{table}[H]
    \centering
    \begin{tabular}{>{\raggedright\arraybackslash}p{1.5cm}>{\raggedright\arraybackslash}p{5cm}>{\raggedright\arraybackslash}p{9.5cm}}
        \textbf{Symbol} & \textbf{Values} & \textbf{Description} \\
        \toprule
        \rowcolor{gray!20}\multicolumn{3}{c}{\textbf{Observed data}} \\
        \midrule
        $q_{m,y}$ & Time series of observed monthly prices, deflated & See more details in the \hyperlink{methods2_data}{Data} section \\
        $z_{k,y}$ & Time series of observed variables, considered to become explanatory variables & \\
        \addlinespace[0.3em]
        \midrule
        \rowcolor{gray!20}
        \multicolumn{3}{c}{\textbf{Variables in the model (in relative annual change units)}} \\
        \midrule
        $p$ & $P = (p_{1}, (p_{2},…,(p_{T})$ & Model input, observations in training set. Time series of relative annual price change \\
        $x$ & $X^{o}_{t} = (x_{t,1},x_{t,2},…,x_{t,K})^{o}$ \\
        & $X^{o}_{t} = (x_{1,k},x_{2,k},…,x_{T,k})^{o}$ & Model input, observations in training set (no TBATS) \\
        $\hat{p}_{y,m}$ & - & Relative price change (annual) to forecast, 1 observation/year \\
        $\hat{p}_{y,m^{ts}}$ & - & Relative price change (annual) to forecast, using time series models, 12 observations/year \\
        \addlinespace[0.3em]
        \midrule
        \rowcolor{gray!20}
        \multicolumn{3}{c}{\textbf{Indices}} \\
        \midrule
        $y$ & $y=1,2,…,Y$ & Years observed ($y\geq$1961) \\
        $k$ & $k=1,2,…,K$ & Number of features in model \\
        $m$ & $m=1,2,…,12$ & Month, fixed in variables of yearly frequency \\
        $h$ & $h=1,2,…,H$ & Lag/Forecasting horizon (1$\leq H\leq $12), in monthly units \\
        $t$ & $t=1,2,…,T$, $1 \leq\hat{t}^{id}\leq T$ & Observations in training set (XML) \\
        $t^{ts}$ & $t^{ts}=1,2,…,T^{ts}$ & Observations in training set (TS) \\
        $y^{f}$ & $y_{1}+45\leq y^{f}\leq Y+1$ & A year in XML testing set, one-step ahead forecast ($f$) \\
        \bottomrule
    \end{tabular}
    \caption{List of variables and indices used in the paper.}
    \label{tab_var_list}
\end{table}
\begin{table}[H]
    \centering
    \begin{tabular}{>{\raggedright\arraybackslash}p{2.5cm}>{\raggedright\arraybackslash}p{3.5cm}>{\raggedright\arraybackslash}p{2cm}>{\raggedright\arraybackslash}p{2.5cm}>{\raggedright\arraybackslash}p{2.5cm}}
        \toprule
        \textbf{Data} & \textbf{Units} & \textbf{AC} & \textbf{Time-range} & \textbf{Source} \\
        \midrule
        \rowcolor{gray!20}
        \multicolumn{5}{c}{\textbf{Final Data}} \\
        \midrule
        Production & \% change /year & M,S,W & 1961 - 2022 & Regional, local \\
        Yield & \% change /year & M,S,W & 1961 - 2022 & Regional, local \\
        Stocks & \% change /year & M,S,W & 1961 - 2022 & Regional, local \\
        Price & \% change /year &  & 01/1961 - 06/2024 & Global \\
        \addlinespace[0.3em]
        \midrule
        \rowcolor{gray!20}
        \multicolumn{5}{c}{\textbf{Initial Information}} \\
        \midrule
        Price & Nominal USD / mt &  & 01/1960 - & World Bank, \\
        Price indices & USD (2010 = 100) &  & 06/2024  & Pink Sheet (2024) \\
        Production & tonnes / year & M,S,W & 1961 - 2022 & FAO STAT (2024) \\
        Yield & hg / ha & M,S,W & 1961 - 2022 & FAO STAT (2024) \\
        Beginning stocks & 1000 mt / year & M,S**,W & 1961 - 2022 & PSD, USDA (2024) \\
        \bottomrule
    \end{tabular}
        \begin{tablenotes}[flushleft]\footnotesize
\item[*] * Commodities: Maize (M), Soybean (S), Wheat (W)
\item[**]** Beginning stocks information for soybean starts in 1963
    \end{tablenotes}
    \caption{Variable description and data sources included in the article.}
    \label{data_sources}
\end{table}
\section{Data adaptation - Model output and input}\label{appendix_deflation}
Defining $p_{m,y}^{n}$ as the nominal price for a given month $m$ within year $y$, $p_{m,y}^{d}$ as the deflated prices, and $In_{m,y}$ as the price index for the same period, with the base year set as 2010 ($In_{m,2010} \approx 100$), the deflation is executed according to the formula:

\begin{equation}\label{eq_1}
p_{m,y}^{d}= \frac{p_{m,y}^{n} \times In_{m,2010}}{In_{m,y}}
\end{equation}

Changes in the global supply of AC's are linked to the local Market Year of each area, as the timing of harvesting and market availability in different regions directly impacts the overall global supply chain and commodity prices \citep{fasusda}.
Accordingly, the dependent variable in the analysis was defined as the proportion of price change relative to the corresponding month ($m$) of the preceding year, as expressed by Eq.\ref{eq_2}:

\begin{equation} \label{eq_2}
p_{m,y}=\frac{p_{m,y}^{d}-p_{m,y-1}^{d}}{p_{m,y-1}^{d}}
\end{equation}
for any given year $y$.

Let $q_{k,y}$ represent the reported production, yield, or stocks in a geographical unit $k$ ($k$=1, …, $K$). Following the methodology outlined in the relative price change function, we further processed the national and regional data to calculate relative annual changes, as shown in Eq.\ref{Eq3}:
\begin{equation} \label{Eq3}
x_{k,y}=\frac{q_{k,y}-q_{k,y-1}}{q_{k,y-1}}
\end{equation}
\end{subappendices}

\begin{subappendices}
\section[Appendix B]{Model Stages}\label{appendix_calculation}
\subsection{Inclusion of possible multicollinearities in the dataset}\label{multicollinearities}
Stage 1 of AGRICAF includes the creation of the dataset used in the retrospective analysis of Stage 2. The dataset is composed of about 100 independent variables, some of them show multicollinearity, meaning that one variable can be linearly predicted from the others with a substantial degree of accuracy. This correlation can present challenges in certain types of models, by inflating the variance of the estimated coefficients, leading to less reliable and less interpretable results \citep{dormann2013collinearity}.

However, this challenge is not true for all types of models. In machine tree-based learning models such as CART, Random Forest, Gradient Boosting, and XGBoost (with a tree booster) which are used in AGRICAF, the presence of multicollinearity is generally less problematic. These models are non-linear and ensemble-based, meaning they build predictions by aggregating the outcomes of multiple decision trees. The use of decision trees in these models allows them to manage multicollinearity by selecting variables that contribute the most to reducing the prediction error at each split \citep{breiman2001machinelearnig, hastie2009elements}. Therefore, even with correlated features, these models can perform robustly without the need for extensive pre-screening for multicollinearity.

The AGRICAF methodology involves the application of three types of linear models: the standard multivariate linear model (LM), Generalised Additive Model (GAM) and XGBoost with a linear booster. In each one of these models, the solution for the presence of multicollinearity is different, as detailed in the \hyperlink{03_methods}{Methods} section.

\subsection{Feature screening and model weighting}\label{screening 2}
\begin{enumerate}
    \item \textbf{Merge Records and Calculate Model Weights}: The 'records' matrix contains observed prices and prices predicted by the model using leave-one-out cross-validation. The records are grouped by year and model, with each model having at least four versions of input data. A weight function is defined to calculate the weight of each model relative to its error. The function aggregates errors of each model, selects the two options with the lowest error, and calculates the weight of each model based on its error relative to the total error.
    \item \textbf{Calculate Relative Importance}: The 'rank' matrix contains the relative influence of each variable in the model. The importance values are standardized, and negative or infinite values are replaced.
    The relative importance values are scaled within the range of 0 to 1. Model weights are applied to the importance values, and the mean importance for each variable is calculated.
    \item \textbf{Build New Dataset}: The dataset used S2 is loaded and filtered.
\end{enumerate}
\end{subappendices}

\newpage
\begin{subappendices}
\section[Appendix C]{Model Performance}\label{appendix_results}
\input{}

\begin{figure}
    \centering
        \includegraphics[width=\textwidth]{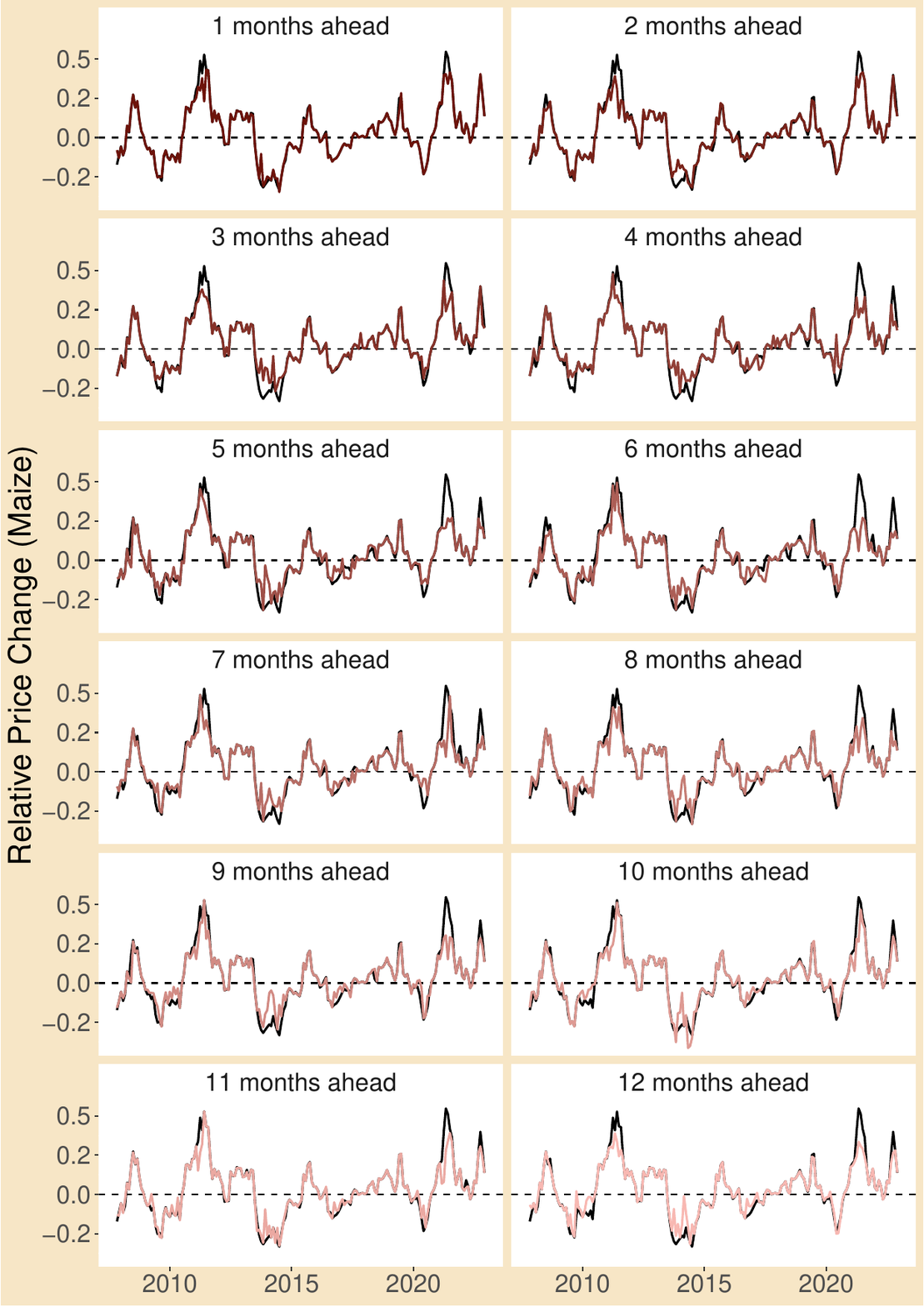}
    \caption{}
    \label{pred_obs_maize}
\end{figure}

\begin{figure}
    \centering
        \includegraphics[width=\textwidth]{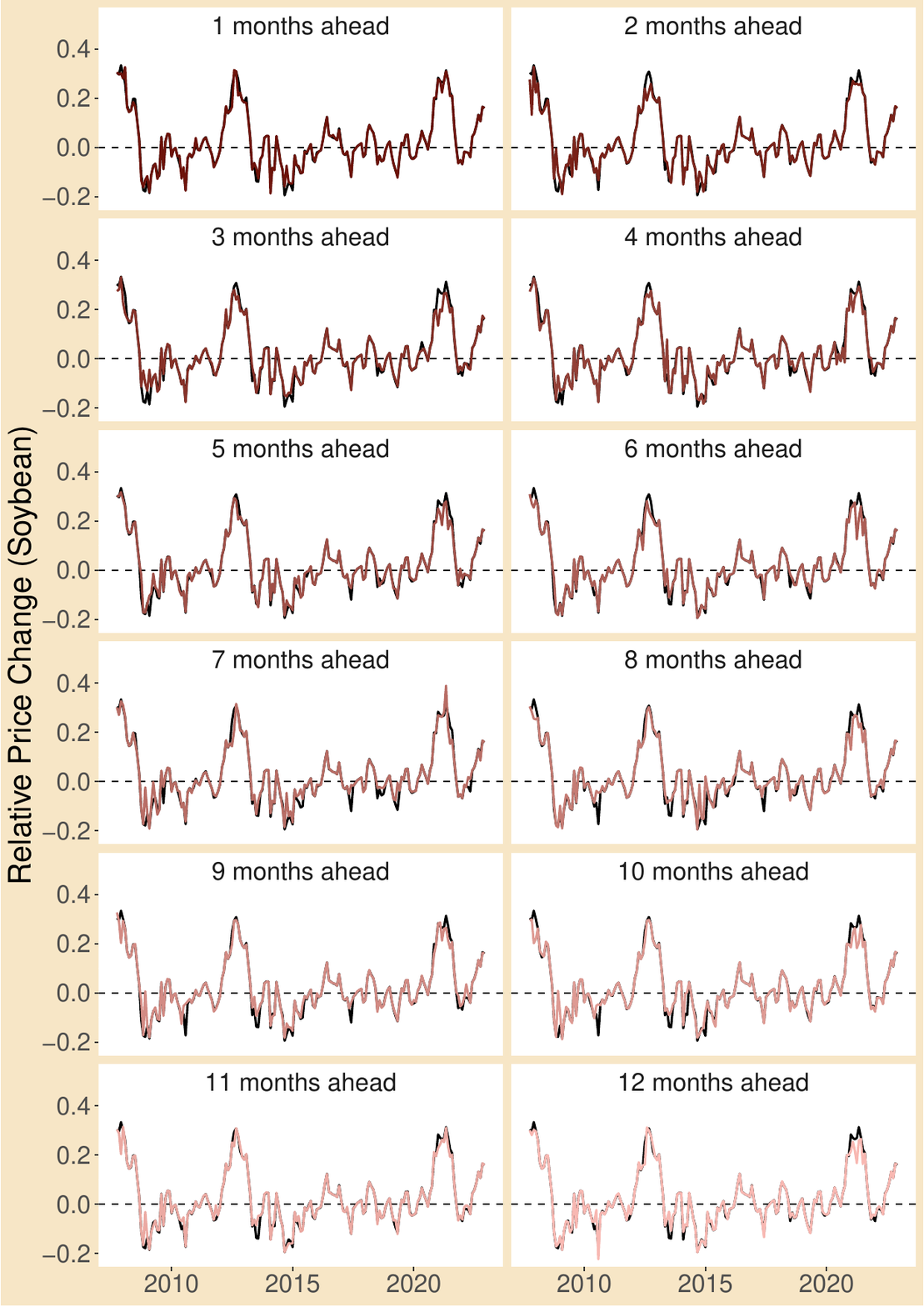}
    \caption{}
    \label{pred_obs_soybean}
\end{figure}

\begin{figure}
    \centering
        \includegraphics[width=\textwidth]{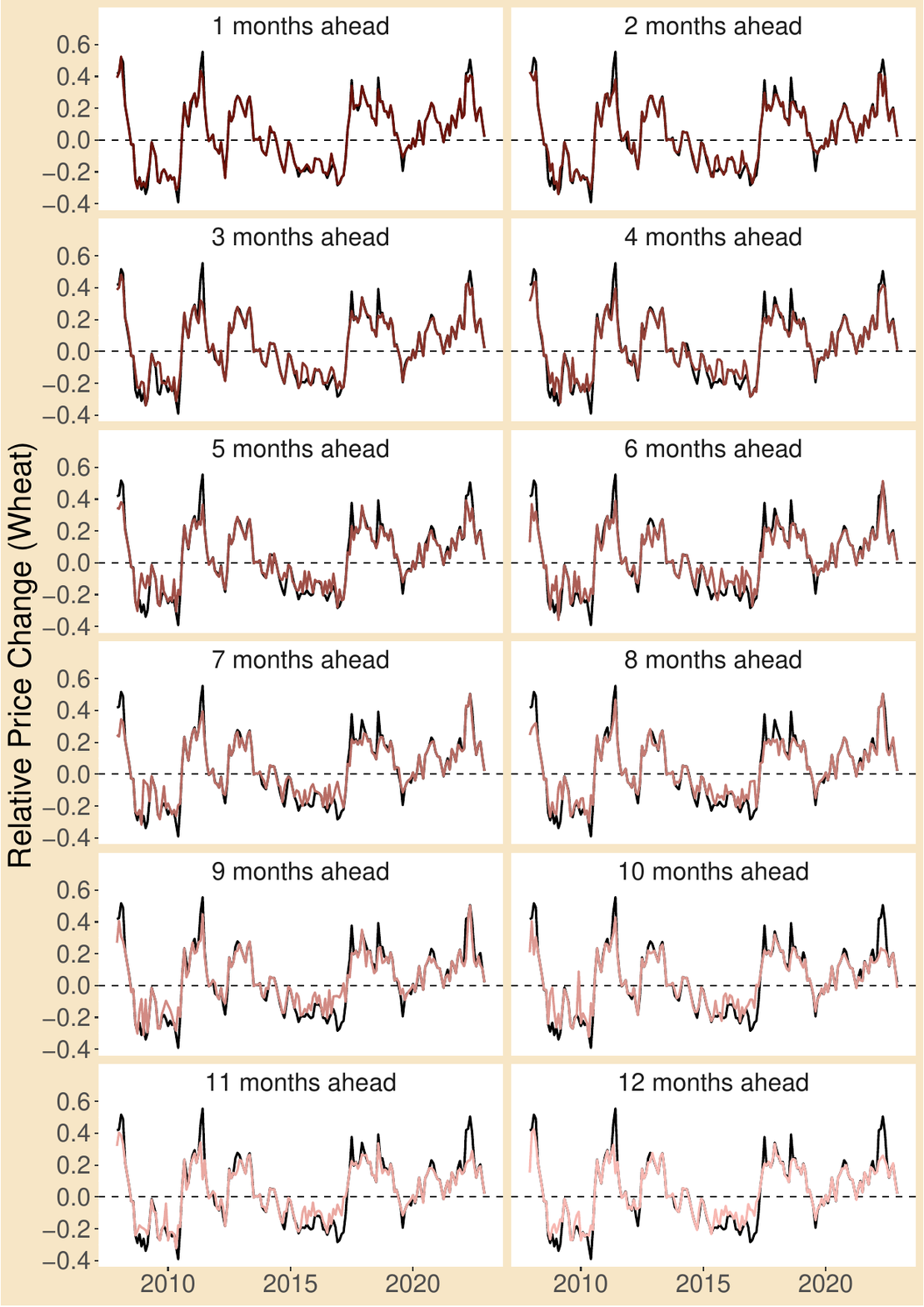}
    \caption{}
    \label{pred_obs_wheat}
\end{figure}

Fig.~\ref{pred_obs_maize}, ~\ref{pred_obs_soybean}, and ~\ref{pred_obs_wheat} present a comparative analysis of the observed and predicted relative annual price changes (monthly prices from December 2007 to 2024) for maize, soybean, and wheat, respectively, corresponding to forecasts made 1 to 12 months ahead.

Tab.~\ref{mae_maize},~\ref{mae_soybean} and ~\ref{mae_wheat} show the MAE values obtained by the favourable forecasting strategies across various months (rows), forecast horizons (columns) and AC's. 
Similarly, Tab.~\ref{ra_maize},~\ref{ra_soybean} and ~\ref{ra_wheat} show highest Relative Advantage (RA) in predicting price changes for each month, forecast horizon and AC. 

\begin{table}
\centering
\caption{Maize}
\centering
\begin{tabular}[t]{rrrrrrrrrrrrr}
\toprule
\multicolumn{1}{c}{} & \multicolumn{12}{c}{MAE (by forecast horizon)} \\
\cmidrule(l{3pt}r{3pt}){2-13}
month & 1 & 2 & 3 & 4 & 5 & 6 & 7 & 8 & 9 & 10 & 11 & 12\\
\midrule
\cellcolor{gray!10}{1} & \cellcolor{gray!10}{0.005} & \cellcolor{gray!10}{0.009} & \cellcolor{gray!10}{0.017} & \cellcolor{gray!10}{0.017} & \cellcolor{gray!10}{0.015} & \cellcolor{gray!10}{0.014} & \cellcolor{gray!10}{0.013} & \cellcolor{gray!10}{0.019} & \cellcolor{gray!10}{0.019} & \cellcolor{gray!10}{0.017} & \cellcolor{gray!10}{0.005} & \cellcolor{gray!10}{0.004}\\
2 & 0.007 & 0.007 & 0.007 & 0.015 & 0.018 & 0.018 & 0.021 & 0.040 & 0.031 & 0.033 & 0.011 & 0.011\\
\cellcolor{gray!10}{3} & \cellcolor{gray!10}{0.002} & \cellcolor{gray!10}{0.014} & \cellcolor{gray!10}{0.012} & \cellcolor{gray!10}{0.010} & \cellcolor{gray!10}{0.029} & \cellcolor{gray!10}{0.022} & \cellcolor{gray!10}{0.020} & \cellcolor{gray!10}{0.031} & \cellcolor{gray!10}{0.034} & \cellcolor{gray!10}{0.030} & \cellcolor{gray!10}{0.035} & \cellcolor{gray!10}{0.036}\\
4 & 0.016 & 0.019 & 0.024 & 0.024 & 0.035 & 0.028 & 0.024 & 0.020 & 0.039 & 0.035 & 0.049 & 0.040\\
\cellcolor{gray!10}{5} & \cellcolor{gray!10}{0.014} & \cellcolor{gray!10}{0.028} & \cellcolor{gray!10}{0.042} & \cellcolor{gray!10}{0.046} & \cellcolor{gray!10}{0.038} & \cellcolor{gray!10}{0.059} & \cellcolor{gray!10}{0.048} & \cellcolor{gray!10}{0.042} & \cellcolor{gray!10}{0.038} & \cellcolor{gray!10}{0.041} & \cellcolor{gray!10}{0.035} & \cellcolor{gray!10}{0.028}\\
\addlinespace
6 & 0.043 & 0.028 & 0.040 & 0.064 & 0.063 & 0.068 & 0.059 & 0.060 & 0.058 & 0.037 & 0.032 & 0.037\\
\cellcolor{gray!10}{7} & \cellcolor{gray!10}{0.006} & \cellcolor{gray!10}{0.035} & \cellcolor{gray!10}{0.038} & \cellcolor{gray!10}{0.050} & \cellcolor{gray!10}{0.047} & \cellcolor{gray!10}{0.055} & \cellcolor{gray!10}{0.048} & \cellcolor{gray!10}{0.008} & \cellcolor{gray!10}{0.040} & \cellcolor{gray!10}{0.007} & \cellcolor{gray!10}{0.008} & \cellcolor{gray!10}{0.045}\\
8 & 0.002 & 0.023 & 0.023 & 0.027 & 0.035 & 0.035 & 0.035 & 0.028 & 0.020 & 0.012 & 0.013 & 0.022\\
\cellcolor{gray!10}{9} & \cellcolor{gray!10}{0.006} & \cellcolor{gray!10}{0.008} & \cellcolor{gray!10}{0.018} & \cellcolor{gray!10}{0.022} & \cellcolor{gray!10}{0.022} & \cellcolor{gray!10}{0.017} & \cellcolor{gray!10}{0.016} & \cellcolor{gray!10}{0.025} & \cellcolor{gray!10}{0.018} & \cellcolor{gray!10}{0.016} & \cellcolor{gray!10}{0.003} & \cellcolor{gray!10}{0.011}\\
10 & 0.017 & 0.012 & 0.018 & 0.041 & 0.038 & 0.039 & 0.042 & 0.041 & 0.021 & 0.018 & 0.018 & 0.041\\
\addlinespace
\cellcolor{gray!10}{11} & \cellcolor{gray!10}{0.009} & \cellcolor{gray!10}{0.015} & \cellcolor{gray!10}{0.023} & \cellcolor{gray!10}{0.025} & \cellcolor{gray!10}{0.014} & \cellcolor{gray!10}{0.014} & \cellcolor{gray!10}{0.013} & \cellcolor{gray!10}{0.012} & \cellcolor{gray!10}{0.011} & \cellcolor{gray!10}{0.010} & \cellcolor{gray!10}{0.008} & \cellcolor{gray!10}{0.014}\\
12 & 0.004 & 0.017 & 0.016 & 0.017 & 0.024 & 0.032 & 0.029 & 0.028 & 0.016 & 0.021 & 0.017 & 0.015\\
\bottomrule
\end{tabular}
\label{mae_maize}
\end{table}

\begin{table}
\centering
\caption{Soybean}
\centering
\begin{tabular}[t]{rrrrrrrrrrrrr}
\toprule
\multicolumn{1}{c}{} & \multicolumn{12}{c}{MAE (by forecast horizon)} \\
\cmidrule(l{3pt}r{3pt}){2-13}
month & 1 & 2 & 3 & 4 & 5 & 6 & 7 & 8 & 9 & 10 & 11 & 12\\
\midrule
\cellcolor{gray!10}{1} & \cellcolor{gray!10}{0.009} & \cellcolor{gray!10}{0.012} & \cellcolor{gray!10}{0.023} & \cellcolor{gray!10}{0.005} & \cellcolor{gray!10}{0.007} & \cellcolor{gray!10}{0.006} & \cellcolor{gray!10}{0.004} & \cellcolor{gray!10}{0.031} & \cellcolor{gray!10}{0.006} & \cellcolor{gray!10}{0.023} & \cellcolor{gray!10}{0.007} & \cellcolor{gray!10}{0.007}\\
2 & 0.009 & 0.002 & 0.020 & 0.018 & 0.011 & 0.002 & 0.003 & 0.003 & 0.003 & 0.004 & 0.004 & 0.004\\
\cellcolor{gray!10}{3} & \cellcolor{gray!10}{0.004} & \cellcolor{gray!10}{0.003} & \cellcolor{gray!10}{0.007} & \cellcolor{gray!10}{0.009} & \cellcolor{gray!10}{0.010} & \cellcolor{gray!10}{0.003} & \cellcolor{gray!10}{0.004} & \cellcolor{gray!10}{0.005} & \cellcolor{gray!10}{0.009} & \cellcolor{gray!10}{0.006} & \cellcolor{gray!10}{0.002} & \cellcolor{gray!10}{0.005}\\
4 & 0.002 & 0.003 & 0.007 & 0.002 & 0.010 & 0.016 & 0.012 & 0.007 & 0.003 & 0.017 & 0.010 & 0.010\\
\cellcolor{gray!10}{5} & \cellcolor{gray!10}{0.004} & \cellcolor{gray!10}{0.009} & \cellcolor{gray!10}{0.012} & \cellcolor{gray!10}{0.007} & \cellcolor{gray!10}{0.005} & \cellcolor{gray!10}{0.014} & \cellcolor{gray!10}{0.025} & \cellcolor{gray!10}{0.010} & \cellcolor{gray!10}{0.007} & \cellcolor{gray!10}{0.011} & \cellcolor{gray!10}{0.003} & \cellcolor{gray!10}{0.006}\\
\addlinespace
6 & 0.005 & 0.012 & 0.007 & 0.016 & 0.014 & 0.016 & 0.016 & 0.017 & 0.019 & 0.005 & 0.006 & 0.005\\
\cellcolor{gray!10}{7} & \cellcolor{gray!10}{0.014} & \cellcolor{gray!10}{0.004} & \cellcolor{gray!10}{0.007} & \cellcolor{gray!10}{0.007} & \cellcolor{gray!10}{0.006} & \cellcolor{gray!10}{0.014} & \cellcolor{gray!10}{0.018} & \cellcolor{gray!10}{0.017} & \cellcolor{gray!10}{0.013} & \cellcolor{gray!10}{0.012} & \cellcolor{gray!10}{0.010} & \cellcolor{gray!10}{0.023}\\
8 & 0.004 & 0.010 & 0.004 & 0.006 & 0.007 & 0.006 & 0.016 & 0.015 & 0.015 & 0.015 & 0.012 & 0.008\\
\cellcolor{gray!10}{9} & \cellcolor{gray!10}{0.004} & \cellcolor{gray!10}{0.008} & \cellcolor{gray!10}{0.013} & \cellcolor{gray!10}{0.010} & \cellcolor{gray!10}{0.007} & \cellcolor{gray!10}{0.007} & \cellcolor{gray!10}{0.008} & \cellcolor{gray!10}{0.002} & \cellcolor{gray!10}{0.004} & \cellcolor{gray!10}{0.004} & \cellcolor{gray!10}{0.001} & \cellcolor{gray!10}{0.002}\\
10 & 0.013 & 0.009 & 0.006 & 0.016 & 0.016 & 0.011 & 0.013 & 0.015 & 0.010 & 0.001 & 0.001 & 0.002\\
\addlinespace
\cellcolor{gray!10}{11} & \cellcolor{gray!10}{0.003} & \cellcolor{gray!10}{0.025} & \cellcolor{gray!10}{0.014} & \cellcolor{gray!10}{0.006} & \cellcolor{gray!10}{0.002} & \cellcolor{gray!10}{0.004} & \cellcolor{gray!10}{0.004} & \cellcolor{gray!10}{0.003} & \cellcolor{gray!10}{0.008} & \cellcolor{gray!10}{0.002} & \cellcolor{gray!10}{0.009} & \cellcolor{gray!10}{0.008}\\
12 & 0.007 & 0.008 & 0.008 & 0.011 & 0.007 & 0.008 & 0.015 & 0.009 & 0.022 & 0.022 & 0.019 & 0.005\\
\bottomrule
\end{tabular}
\label{mae_soybean}
\end{table}

\begin{table}
\centering
\caption{Wheat}
\centering
\begin{tabular}[t]{rrrrrrrrrrrrr}
\toprule
\multicolumn{1}{c}{} & \multicolumn{12}{c}{MAE (by forecast horizon)} \\
\cmidrule(l{3pt}r{3pt}){2-13}
month & 1 & 2 & 3 & 4 & 5 & 6 & 7 & 8 & 9 & 10 & 11 & 12\\
\midrule
\cellcolor{gray!10}{1} & \cellcolor{gray!10}{0.002} & \cellcolor{gray!10}{0.005} & \cellcolor{gray!10}{0.020} & \cellcolor{gray!10}{0.013} & \cellcolor{gray!10}{0.024} & \cellcolor{gray!10}{0.019} & \cellcolor{gray!10}{0.047} & \cellcolor{gray!10}{0.044} & \cellcolor{gray!10}{0.040} & \cellcolor{gray!10}{0.019} & \cellcolor{gray!10}{0.025} & \cellcolor{gray!10}{0.022}\\
2 & 0.005 & 0.012 & 0.007 & 0.020 & 0.027 & 0.029 & 0.037 & 0.067 & 0.067 & 0.058 & 0.031 & 0.028\\
\cellcolor{gray!10}{3} & \cellcolor{gray!10}{0.017} & \cellcolor{gray!10}{0.022} & \cellcolor{gray!10}{0.008} & \cellcolor{gray!10}{0.015} & \cellcolor{gray!10}{0.028} & \cellcolor{gray!10}{0.038} & \cellcolor{gray!10}{0.030} & \cellcolor{gray!10}{0.033} & \cellcolor{gray!10}{0.057} & \cellcolor{gray!10}{0.053} & \cellcolor{gray!10}{0.049} & \cellcolor{gray!10}{0.047}\\
4 & 0.006 & 0.006 & 0.019 & 0.030 & 0.028 & 0.011 & 0.008 & 0.010 & 0.014 & 0.022 & 0.030 & 0.024\\
\cellcolor{gray!10}{5} & \cellcolor{gray!10}{0.017} & \cellcolor{gray!10}{0.030} & \cellcolor{gray!10}{0.023} & \cellcolor{gray!10}{0.040} & \cellcolor{gray!10}{0.038} & \cellcolor{gray!10}{0.046} & \cellcolor{gray!10}{0.024} & \cellcolor{gray!10}{0.022} & \cellcolor{gray!10}{0.031} & \cellcolor{gray!10}{0.035} & \cellcolor{gray!10}{0.032} & \cellcolor{gray!10}{0.045}\\
\addlinespace
6 & 0.021 & 0.022 & 0.031 & 0.033 & 0.054 & 0.056 & 0.052 & 0.046 & 0.060 & 0.057 & 0.059 & 0.060\\
\cellcolor{gray!10}{7} & \cellcolor{gray!10}{0.006} & \cellcolor{gray!10}{0.018} & \cellcolor{gray!10}{0.026} & \cellcolor{gray!10}{0.021} & \cellcolor{gray!10}{0.016} & \cellcolor{gray!10}{0.015} & \cellcolor{gray!10}{0.016} & \cellcolor{gray!10}{0.020} & \cellcolor{gray!10}{0.023} & \cellcolor{gray!10}{0.022} & \cellcolor{gray!10}{0.022} & \cellcolor{gray!10}{0.024}\\
8 & 0.015 & 0.038 & 0.038 & 0.041 & 0.050 & 0.040 & 0.039 & 0.032 & 0.041 & 0.023 & 0.026 & 0.025\\
\cellcolor{gray!10}{9} & \cellcolor{gray!10}{0.010} & \cellcolor{gray!10}{0.015} & \cellcolor{gray!10}{0.035} & \cellcolor{gray!10}{0.051} & \cellcolor{gray!10}{0.007} & \cellcolor{gray!10}{0.013} & \cellcolor{gray!10}{0.005} & \cellcolor{gray!10}{0.011} & \cellcolor{gray!10}{0.003} & \cellcolor{gray!10}{0.002} & \cellcolor{gray!10}{0.003} & \cellcolor{gray!10}{0.004}\\
10 & 0.006 & 0.012 & 0.011 & 0.031 & 0.038 & 0.043 & 0.017 & 0.021 & 0.042 & 0.028 & 0.050 & 0.019\\
\addlinespace
\cellcolor{gray!10}{11} & \cellcolor{gray!10}{0.002} & \cellcolor{gray!10}{0.008} & \cellcolor{gray!10}{0.003} & \cellcolor{gray!10}{0.010} & \cellcolor{gray!10}{0.022} & \cellcolor{gray!10}{0.043} & \cellcolor{gray!10}{0.033} & \cellcolor{gray!10}{0.012} & \cellcolor{gray!10}{0.030} & \cellcolor{gray!10}{0.041} & \cellcolor{gray!10}{0.030} & \cellcolor{gray!10}{0.036}\\
12 & 0.009 & 0.020 & 0.020 & 0.018 & 0.026 & 0.075 & 0.061 & 0.059 & 0.060 & 0.076 & 0.069 & 0.061\\
\bottomrule
\end{tabular}
\label{mae_wheat}
\end{table}

\begin{table}
\centering
\caption{Maize}
\centering
\begin{tabular}[t]{rrrrrrrrrrrrr}
\toprule
\multicolumn{1}{c}{} & \multicolumn{12}{c}{RA (by forecast horizon)} \\
\cmidrule(l{3pt}r{3pt}){2-13}
month & 1 & 2 & 3 & 4 & 5 & 6 & 7 & 8 & 9 & 10 & 11 & 12\\
\midrule
\cellcolor{gray!10}{1} & \cellcolor{gray!10}{0.94} & \cellcolor{gray!10}{0.87} & \cellcolor{gray!10}{0.75} & \cellcolor{gray!10}{0.77} & \cellcolor{gray!10}{0.72} & \cellcolor{gray!10}{0.76} & \cellcolor{gray!10}{0.79} & \cellcolor{gray!10}{0.73} & \cellcolor{gray!10}{0.72} & \cellcolor{gray!10}{0.78} & \cellcolor{gray!10}{0.93} & \cellcolor{gray!10}{0.94}\\
2 & 0.90 & 0.87 & 0.90 & 0.74 & 0.75 & 0.68 & 0.67 & 0.56 & 0.57 & 0.54 & 0.85 & 0.85\\
\cellcolor{gray!10}{3} & \cellcolor{gray!10}{0.98} & \cellcolor{gray!10}{0.79} & \cellcolor{gray!10}{0.84} & \cellcolor{gray!10}{0.79} & \cellcolor{gray!10}{0.71} & \cellcolor{gray!10}{0.66} & \cellcolor{gray!10}{0.74} & \cellcolor{gray!10}{0.58} & \cellcolor{gray!10}{0.58} & \cellcolor{gray!10}{0.65} & \cellcolor{gray!10}{0.56} & \cellcolor{gray!10}{0.51}\\
4 & 0.74 & 0.77 & 0.70 & 0.80 & 0.67 & 0.74 & 0.71 & 0.79 & 0.67 & 0.60 & 0.51 & 0.60\\
\cellcolor{gray!10}{5} & \cellcolor{gray!10}{0.82} & \cellcolor{gray!10}{0.72} & \cellcolor{gray!10}{0.58} & \cellcolor{gray!10}{0.58} & \cellcolor{gray!10}{0.57} & \cellcolor{gray!10}{0.51} & \cellcolor{gray!10}{0.49} & \cellcolor{gray!10}{0.57} & \cellcolor{gray!10}{0.64} & \cellcolor{gray!10}{0.58} & \cellcolor{gray!10}{0.63} & \cellcolor{gray!10}{0.70}\\
\addlinespace
6 & 0.77 & 0.74 & 0.69 & 0.59 & 0.59 & 0.57 & 0.56 & 0.57 & 0.56 & 0.76 & 0.73 & 0.68\\
\cellcolor{gray!10}{7} & \cellcolor{gray!10}{0.96} & \cellcolor{gray!10}{0.71} & \cellcolor{gray!10}{0.73} & \cellcolor{gray!10}{0.65} & \cellcolor{gray!10}{0.67} & \cellcolor{gray!10}{0.66} & \cellcolor{gray!10}{0.64} & \cellcolor{gray!10}{0.91} & \cellcolor{gray!10}{0.67} & \cellcolor{gray!10}{0.95} & \cellcolor{gray!10}{0.94} & \cellcolor{gray!10}{0.66}\\
8 & 0.99 & 0.76 & 0.78 & 0.76 & 0.71 & 0.70 & 0.67 & 0.73 & 0.81 & 0.88 & 0.89 & 0.75\\
\cellcolor{gray!10}{9} & \cellcolor{gray!10}{0.95} & \cellcolor{gray!10}{0.93} & \cellcolor{gray!10}{0.79} & \cellcolor{gray!10}{0.73} & \cellcolor{gray!10}{0.73} & \cellcolor{gray!10}{0.85} & \cellcolor{gray!10}{0.85} & \cellcolor{gray!10}{0.69} & \cellcolor{gray!10}{0.80} & \cellcolor{gray!10}{0.85} & \cellcolor{gray!10}{0.96} & \cellcolor{gray!10}{0.88}\\
10 & 0.82 & 0.86 & 0.71 & 0.57 & 0.59 & 0.54 & 0.57 & 0.55 & 0.77 & 0.79 & 0.80 & 0.57\\
\addlinespace
\cellcolor{gray!10}{11} & \cellcolor{gray!10}{0.84} & \cellcolor{gray!10}{0.79} & \cellcolor{gray!10}{0.69} & \cellcolor{gray!10}{0.67} & \cellcolor{gray!10}{0.82} & \cellcolor{gray!10}{0.79} & \cellcolor{gray!10}{0.82} & \cellcolor{gray!10}{0.81} & \cellcolor{gray!10}{0.88} & \cellcolor{gray!10}{0.87} & \cellcolor{gray!10}{0.89} & \cellcolor{gray!10}{0.78}\\
12 & 0.96 & 0.74 & 0.73 & 0.66 & 0.62 & 0.62 & 0.66 & 0.68 & 0.77 & 0.73 & 0.69 & 0.78\\
\bottomrule
\end{tabular}
\label{ra_maize}
\end{table}

\begin{table}
\centering
\caption{Soybean}
\centering
\begin{tabular}[t]{rrrrrrrrrrrrr}
\toprule
\multicolumn{1}{c}{} & \multicolumn{12}{c}{RA (by forecast horizon)} \\
\cmidrule(l{3pt}r{3pt}){2-13}
month & 1 & 2 & 3 & 4 & 5 & 6 & 7 & 8 & 9 & 10 & 11 & 12\\
\midrule
\cellcolor{gray!10}{1} & \cellcolor{gray!10}{0.91} & \cellcolor{gray!10}{0.81} & \cellcolor{gray!10}{0.64} & \cellcolor{gray!10}{0.93} & \cellcolor{gray!10}{0.90} & \cellcolor{gray!10}{0.92} & \cellcolor{gray!10}{0.96} & \cellcolor{gray!10}{0.55} & \cellcolor{gray!10}{0.93} & \cellcolor{gray!10}{0.68} & \cellcolor{gray!10}{0.89} & \cellcolor{gray!10}{0.86}\\
2 & 0.85 & 0.98 & 0.71 & 0.75 & 0.82 & 0.97 & 0.96 & 0.96 & 0.96 & 0.94 & 0.94 & 0.93\\
\cellcolor{gray!10}{3} & \cellcolor{gray!10}{0.90} & \cellcolor{gray!10}{0.95} & \cellcolor{gray!10}{0.82} & \cellcolor{gray!10}{0.84} & \cellcolor{gray!10}{0.81} & \cellcolor{gray!10}{0.94} & \cellcolor{gray!10}{0.94} & \cellcolor{gray!10}{0.92} & \cellcolor{gray!10}{0.81} & \cellcolor{gray!10}{0.80} & \cellcolor{gray!10}{0.95} & \cellcolor{gray!10}{0.89}\\
4 & 0.98 & 0.96 & 0.89 & 0.97 & 0.79 & 0.72 & 0.78 & 0.89 & 0.95 & 0.71 & 0.85 & 0.76\\
\cellcolor{gray!10}{5} & \cellcolor{gray!10}{0.93} & \cellcolor{gray!10}{0.84} & \cellcolor{gray!10}{0.82} & \cellcolor{gray!10}{0.88} & \cellcolor{gray!10}{0.91} & \cellcolor{gray!10}{0.74} & \cellcolor{gray!10}{0.64} & \cellcolor{gray!10}{0.78} & \cellcolor{gray!10}{0.87} & \cellcolor{gray!10}{0.76} & \cellcolor{gray!10}{0.97} & \cellcolor{gray!10}{0.88}\\
\addlinespace
6 & 0.92 & 0.80 & 0.89 & 0.68 & 0.78 & 0.79 & 0.73 & 0.73 & 0.73 & 0.91 & 0.89 & 0.93\\
\cellcolor{gray!10}{7} & \cellcolor{gray!10}{0.77} & \cellcolor{gray!10}{0.94} & \cellcolor{gray!10}{0.88} & \cellcolor{gray!10}{0.87} & \cellcolor{gray!10}{0.89} & \cellcolor{gray!10}{0.77} & \cellcolor{gray!10}{0.70} & \cellcolor{gray!10}{0.74} & \cellcolor{gray!10}{0.81} & \cellcolor{gray!10}{0.82} & \cellcolor{gray!10}{0.85} & \cellcolor{gray!10}{0.65}\\
8 & 0.95 & 0.76 & 0.93 & 0.91 & 0.89 & 0.88 & 0.75 & 0.72 & 0.72 & 0.72 & 0.77 & 0.89\\
\cellcolor{gray!10}{9} & \cellcolor{gray!10}{0.94} & \cellcolor{gray!10}{0.81} & \cellcolor{gray!10}{0.79} & \cellcolor{gray!10}{0.82} & \cellcolor{gray!10}{0.89} & \cellcolor{gray!10}{0.84} & \cellcolor{gray!10}{0.83} & \cellcolor{gray!10}{0.97} & \cellcolor{gray!10}{0.95} & \cellcolor{gray!10}{0.95} & \cellcolor{gray!10}{0.98} & \cellcolor{gray!10}{0.97}\\
10 & 0.77 & 0.84 & 0.92 & 0.77 & 0.75 & 0.80 & 0.74 & 0.76 & 0.82 & 0.98 & 0.98 & 0.98\\
\addlinespace
\cellcolor{gray!10}{11} & \cellcolor{gray!10}{0.96} & \cellcolor{gray!10}{0.64} & \cellcolor{gray!10}{0.75} & \cellcolor{gray!10}{0.87} & \cellcolor{gray!10}{0.97} & \cellcolor{gray!10}{0.94} & \cellcolor{gray!10}{0.94} & \cellcolor{gray!10}{0.97} & \cellcolor{gray!10}{0.87} & \cellcolor{gray!10}{0.98} & \cellcolor{gray!10}{0.87} & \cellcolor{gray!10}{0.90}\\
12 & 0.89 & 0.85 & 0.85 & 0.84 & 0.88 & 0.85 & 0.71 & 0.85 & 0.63 & 0.58 & 0.64 & 0.93\\
\bottomrule
\end{tabular}
\label{ra_soybean}
\end{table}

\begin{table}
\centering
\caption{Wheat}
\centering
\begin{tabular}[t]{rrrrrrrrrrrrr}
\toprule
\multicolumn{1}{c}{} & \multicolumn{12}{c}{RA (by forecast horizon)} \\
\cmidrule(l{3pt}r{3pt}){2-13}
month & 1 & 2 & 3 & 4 & 5 & 6 & 7 & 8 & 9 & 10 & 11 & 12\\
\midrule
\cellcolor{gray!10}{1} & \cellcolor{gray!10}{0.99} & \cellcolor{gray!10}{0.95} & \cellcolor{gray!10}{0.83} & \cellcolor{gray!10}{0.87} & \cellcolor{gray!10}{0.78} & \cellcolor{gray!10}{0.77} & \cellcolor{gray!10}{0.62} & \cellcolor{gray!10}{0.64} & \cellcolor{gray!10}{0.68} & \cellcolor{gray!10}{0.78} & \cellcolor{gray!10}{0.76} & \cellcolor{gray!10}{0.76}\\
2 & 0.95 & 0.89 & 0.95 & 0.80 & 0.78 & 0.74 & 0.72 & 0.53 & 0.54 & 0.56 & 0.75 & 0.77\\
\cellcolor{gray!10}{3} & \cellcolor{gray!10}{0.85} & \cellcolor{gray!10}{0.82} & \cellcolor{gray!10}{0.95} & \cellcolor{gray!10}{0.90} & \cellcolor{gray!10}{0.72} & \cellcolor{gray!10}{0.70} & \cellcolor{gray!10}{0.68} & \cellcolor{gray!10}{0.66} & \cellcolor{gray!10}{0.54} & \cellcolor{gray!10}{0.58} & \cellcolor{gray!10}{0.61} & \cellcolor{gray!10}{0.64}\\
4 & 0.94 & 0.96 & 0.82 & 0.78 & 0.74 & 0.89 & 0.91 & 0.85 & 0.83 & 0.72 & 0.69 & 0.72\\
\cellcolor{gray!10}{5} & \cellcolor{gray!10}{0.86} & \cellcolor{gray!10}{0.71} & \cellcolor{gray!10}{0.77} & \cellcolor{gray!10}{0.74} & \cellcolor{gray!10}{0.70} & \cellcolor{gray!10}{0.64} & \cellcolor{gray!10}{0.80} & \cellcolor{gray!10}{0.79} & \cellcolor{gray!10}{0.74} & \cellcolor{gray!10}{0.65} & \cellcolor{gray!10}{0.67} & \cellcolor{gray!10}{0.64}\\
\addlinespace
6 & 0.80 & 0.78 & 0.67 & 0.72 & 0.62 & 0.62 & 0.65 & 0.65 & 0.61 & 0.57 & 0.45 & 0.57\\
\cellcolor{gray!10}{7} & \cellcolor{gray!10}{0.94} & \cellcolor{gray!10}{0.81} & \cellcolor{gray!10}{0.73} & \cellcolor{gray!10}{0.73} & \cellcolor{gray!10}{0.77} & \cellcolor{gray!10}{0.77} & \cellcolor{gray!10}{0.77} & \cellcolor{gray!10}{0.72} & \cellcolor{gray!10}{0.72} & \cellcolor{gray!10}{0.73} & \cellcolor{gray!10}{0.73} & \cellcolor{gray!10}{0.70}\\
8 & 0.83 & 0.64 & 0.65 & 0.62 & 0.58 & 0.58 & 0.60 & 0.65 & 0.58 & 0.69 & 0.73 & 0.71\\
\cellcolor{gray!10}{9} & \cellcolor{gray!10}{0.87} & \cellcolor{gray!10}{0.79} & \cellcolor{gray!10}{0.65} & \cellcolor{gray!10}{0.57} & \cellcolor{gray!10}{0.91} & \cellcolor{gray!10}{0.87} & \cellcolor{gray!10}{0.97} & \cellcolor{gray!10}{0.89} & \cellcolor{gray!10}{0.98} & \cellcolor{gray!10}{0.99} & \cellcolor{gray!10}{0.98} & \cellcolor{gray!10}{0.97}\\
10 & 0.94 & 0.86 & 0.90 & 0.73 & 0.66 & 0.62 & 0.82 & 0.79 & 0.70 & 0.66 & 0.64 & 0.82\\
\addlinespace
\cellcolor{gray!10}{11} & \cellcolor{gray!10}{0.98} & \cellcolor{gray!10}{0.89} & \cellcolor{gray!10}{0.97} & \cellcolor{gray!10}{0.89} & \cellcolor{gray!10}{0.80} & \cellcolor{gray!10}{0.63} & \cellcolor{gray!10}{0.71} & \cellcolor{gray!10}{0.88} & \cellcolor{gray!10}{0.74} & \cellcolor{gray!10}{0.65} & \cellcolor{gray!10}{0.74} & \cellcolor{gray!10}{0.70}\\
12 & 0.93 & 0.84 & 0.84 & 0.86 & 0.71 & 0.49 & 0.58 & 0.59 & 0.56 & 0.47 & 0.57 & 0.55\\
\bottomrule
\end{tabular}
\label{ra_wheat}
\end{table}

\begin{table}[h]
\centering
\begin{tabular}{p{0.5cm} p{1.5cm} p{0.8cm} p{0.9cm} p{6.5cm} p{3.6cm}}
\toprule
AC & Date & $p_{m,y}^d$ & $p_{m,y}$ & Event & Source\\
\midrule
\cellcolor{gray!10}{M} & \cellcolor{gray!10}{Apr-11} & \cellcolor{gray!10}{233.9} & \cellcolor{gray!10}{0.49} & \cellcolor{gray!10}{2010-2011 weather disruptions (Russia drought, US floods) drove maize prices higher} & \cellcolor{gray!10}{\citep{zampieri2016global}}\\
M & May-21 & 254.0 & 0.55 & COVID-19 supply chain disruptions and Chinese demand spike & \citep{espitia2020covid,cariappa2022covid}\\
\cellcolor{gray!10}{M} & \cellcolor{gray!10}{Oct-13} & \cellcolor{gray!10}{213.2} & \cellcolor{gray!10}{-0.30} & \cellcolor{gray!10}{Improved weather in 2013 led to record maize yields, causing prices to drop} & \cellcolor{gray!10}{\citep{schnitkey20132013net}}\\
M & Apr-21 & 237.2 & 0.40 & Continued COVID-19 effects and South American weather issues increased maize prices & \citep{espitia2020covid,cariappa2022covid}\\
\cellcolor{gray!10}{S} & \cellcolor{gray!10}{Jan-15} & \cellcolor{gray!10}{445.1} & \cellcolor{gray!10}{-0.17} & \cellcolor{gray!10}{Global oil price collapse in late 2014 reduced demand for biodiesel} & \cellcolor{gray!10}{\citep{prest2018explanation}}\\
\addlinespace
S & Feb-09 & 415.2 & -0.19 & Global financial crisis in 2008 reduced demand, impacting soybean prices & \citep{sumner2009recent}\\
\cellcolor{gray!10}{S} & \cellcolor{gray!10}{Jun-10} & \cellcolor{gray!10}{401.3} & \cellcolor{gray!10}{-0.10} & \cellcolor{gray!10}{Improved weather led to strong South American harvests, lowering prices} & \cellcolor{gray!10}{\citep{good2011corn}}\\
S & Feb-14 & 445.6 & -0.14 & Record South American harvest and slower Chinese demand lowered soybean prices & \citep{good2013willchina}\\
\cellcolor{gray!10}{W} & \cellcolor{gray!10}{Jun-11} & \cellcolor{gray!10}{245.4} & \cellcolor{gray!10}{0.56} & \cellcolor{gray!10}{2010 Russian drought and export ban significantly reduced wheat supply} & \cellcolor{gray!10}{\citep{zampieri2016global,wegren2011food}}\\
W & May-22 & 372.6 & 0.51 & Russia-Ukraine war caused significant disruptions to wheat exports & \citep{glauber2023russia}\\
\addlinespace
\cellcolor{gray!10}{W} & \cellcolor{gray!10}{Jun-10} & \cellcolor{gray!10}{157.7} & \cellcolor{gray!10}{-0.39} & \cellcolor{gray!10}{Strong wheat production in North America in 2010 reduced prices} & \cellcolor{gray!10}{\citep{taylor20102010outlook}}\\
W & Aug-18 & 278.3 & 0.39 & US-China trade war in 2018 and droughts in Australia caused price increases & \citep{kingwell2020changing}\\
\bottomrule
\end{tabular}
\caption{Major events associated with large forecast errors in 12-month predictions for maize (M), soybean (S), and wheat (W).}
\label{tab_extremes_explained}
\end{table}
\end{subappendices}

\newpage
\begin{subappendices}
\section[Appendix D]{General Interpretation}\label{appendix_interpretation}
Relative importance: These values reflect the mean Root Mean Squared Error (RMSE) increase through the observed period, derived from the exclusion of a feature from the set. 

\begin{figure}[!htb]
    \centering
    \begin{subfigure}[b]{0.49\textwidth}
        \includegraphics[width=\textwidth]{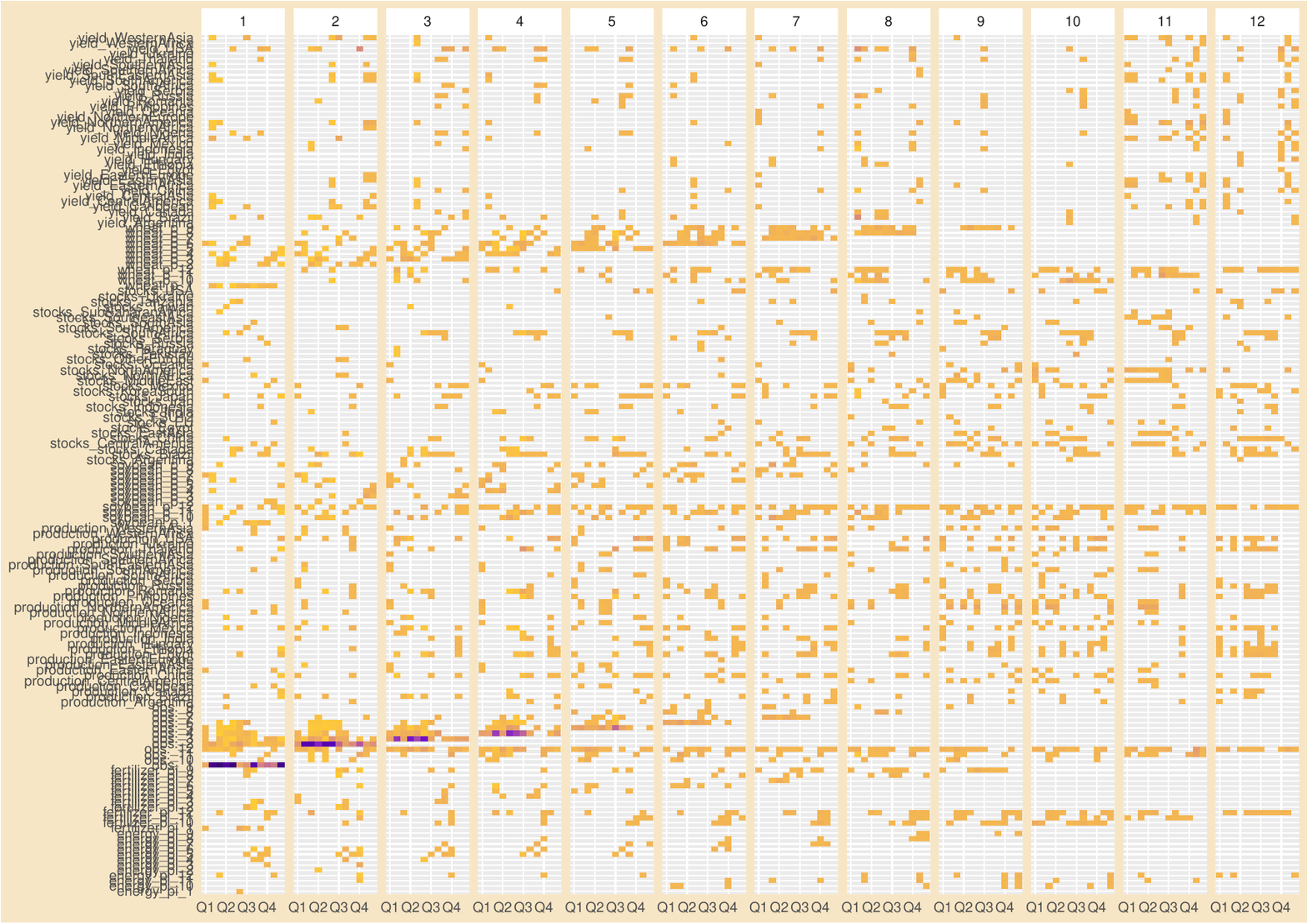}
        \caption{Maize}
        \label{fig:maize}
    \end{subfigure}
    \hfill
    \begin{subfigure}[b]{0.49\textwidth}
        \includegraphics[width=\textwidth]{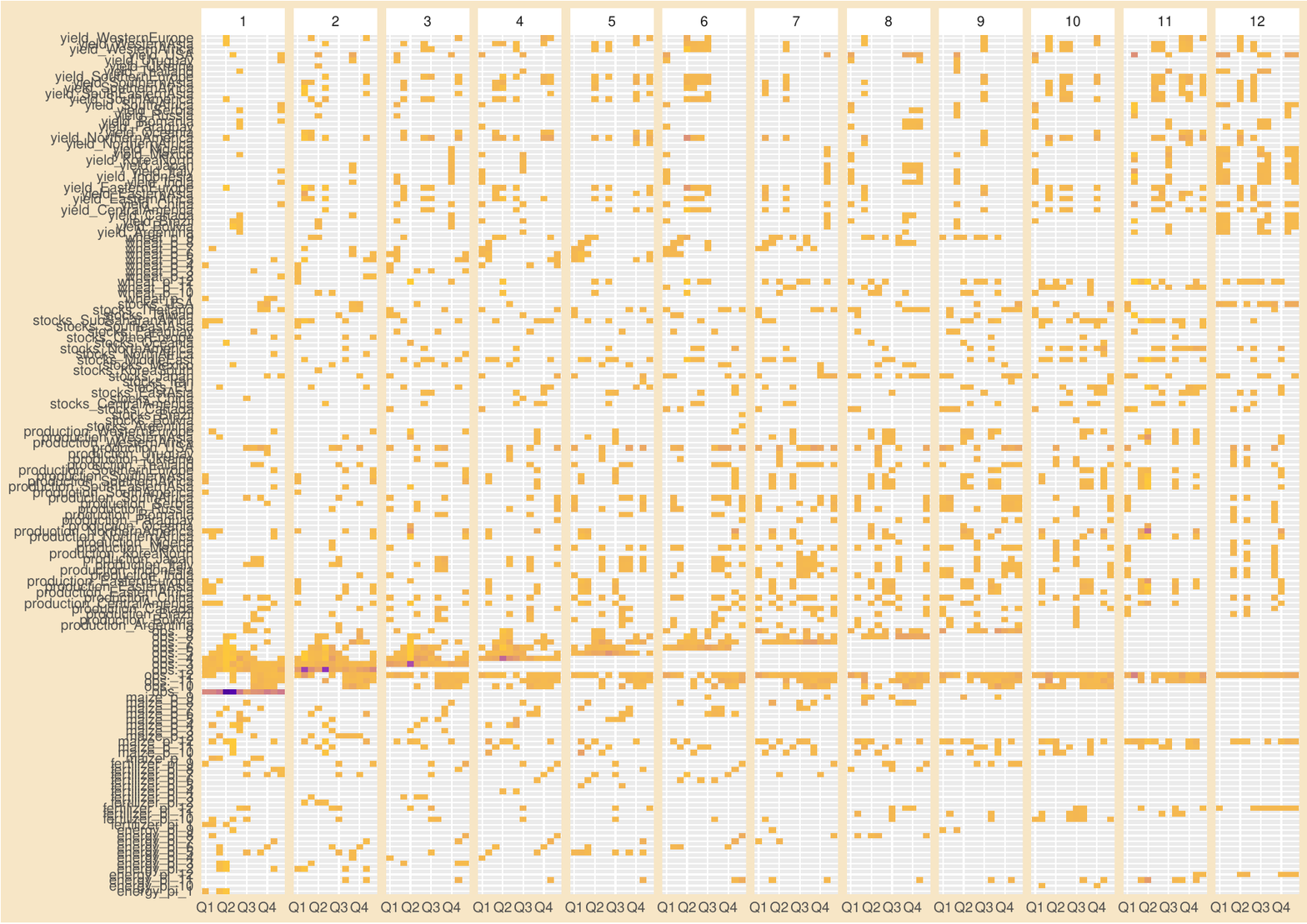}
        \caption{Soybean}
        \label{fig:soybean}
    \end{subfigure}
    
    \begin{subfigure}[b]{0.49\textwidth}
        \includegraphics[width=\textwidth]{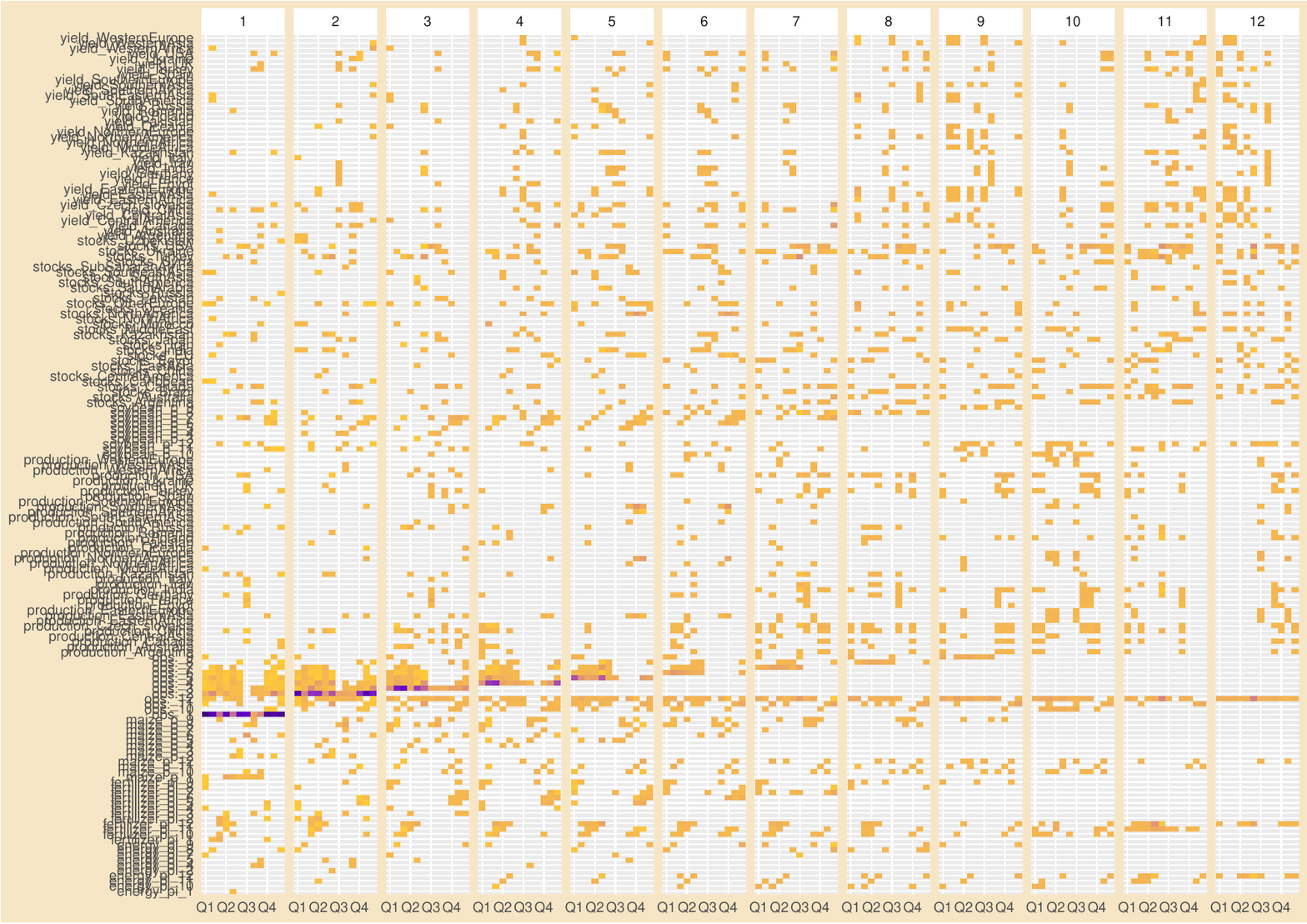}
        \caption{Wheat}
        \label{fig:wheat}
    \end{subfigure}
        \hfill
    \begin{subfigure}[b]{0.49\textwidth}
        \includegraphics[width=\textwidth]{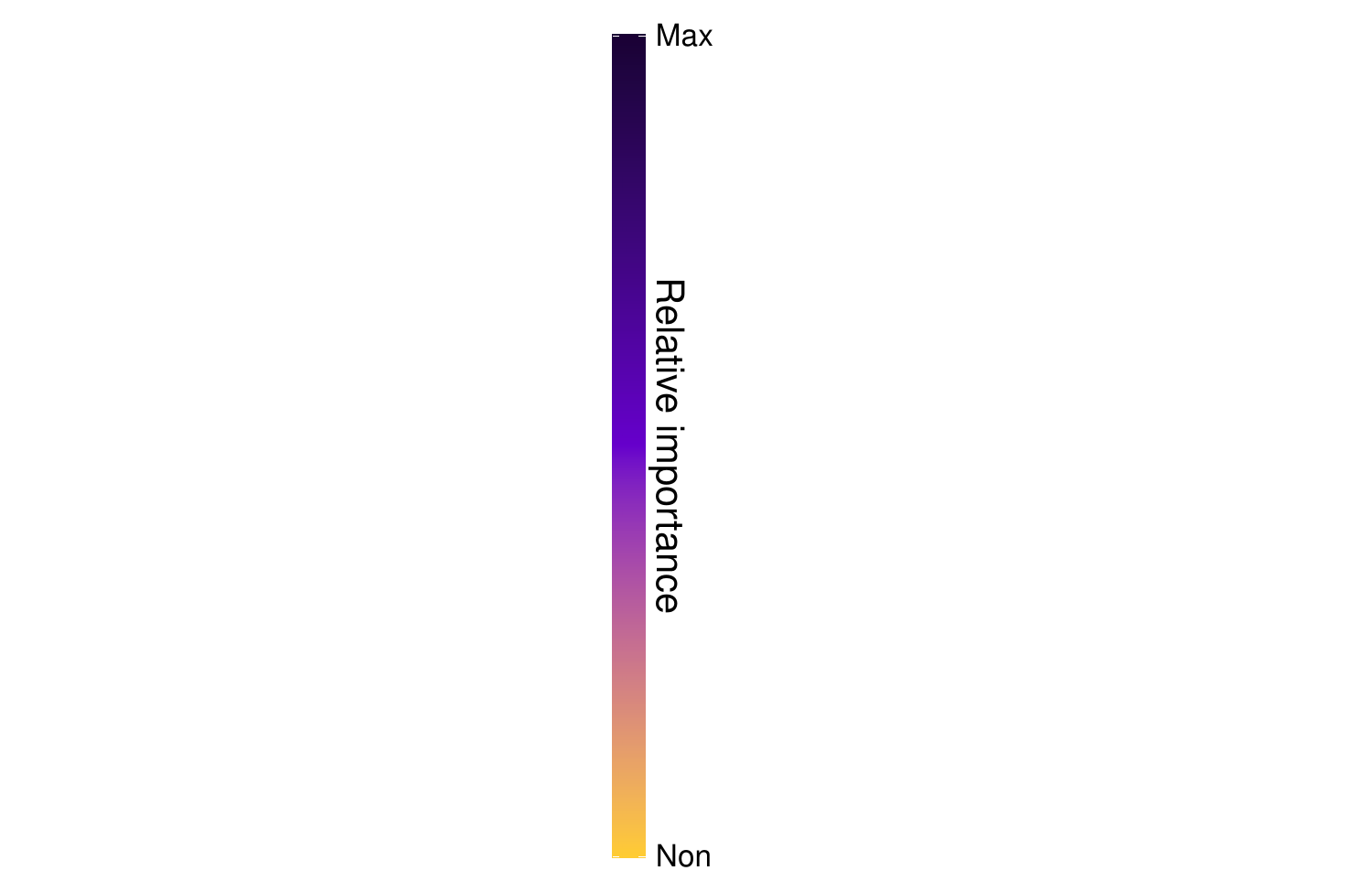}
        \label{fig:legend}
    \end{subfigure}
    \hfill
    \caption{Assessment matrix showing the median relative importance of annual changes in selected features (vertical axis) for monthly price forecasting of maize, soybean, and wheat throughout the year. The months are marked on the horizontal axis as quarters (Q1, Q2, Q3, Q4), with panels representing 1-12 month forecasting horizons. Feature importance is defined as the contribution of each predictor to the RMSE resulting from a random shuffle of the training set. Dark purple indicates predictors with high influence, while orange represents features with a low impact.}
    \label{global_matrix_app}
\end{figure}
\end{subappendices}

\end{document}